\documentclass[superscriptaddress, floatfix, reprint]{revtex4-2}
\setcitestyle{super,open={},close={}}

\usepackage{hyperref, graphicx, xspace}
\hypersetup{hidelinks, allcolors=blue, colorlinks=true}

\usepackage{multirow,booktabs,siunitx,natmove}
\AtBeginDocument{\RenewCommandCopy\qty\SI}
\DeclareSIUnit{\angstrom}{\textup{\AA}}
\DeclareSIUnit{\hartree}{hartree}
\DeclareSIUnit{\bohr}{bohr}

\usepackage[version=4]{mhchem}
\usepackage[capitalize]{cleveref}
\creflabelformat{equation}{#2\textup{#1}#3}

\usepackage[acronym]{glossaries-extra}
\setabbreviationstyle[acronym]{long-postshort-user}
\glsdisablehyper
\newacronym{rdm}{RDM}{reduced density matrix}
\newacronym{ks}{KS}{Kohn-Sham}
\newacronym{mo}{MO}{molecular orbital}
\newacronym{ao}{AO}{atomic orbital}
\newacronym{omo}{OMO}{orthogonal molecular orbital}
\newacronym{ue}{UE}{unsigned error}
\newacronym{mue}{MUE}{mean unsigned error}
\newacronym{re}{RE}{relative error}
\newacronym{tbe}{TBE}{theoretical best estimate}
\newacronym{zpe}{ZPE}{zero-point energy}
\newacronym{mud}{MUD}{mean unsigned deviation}

\newacronym{cc2}{CC2}{second-order coupled cluster}
\newacronym{cc3}{CC3}{third-order coupled cluster}
\newacronym{adc2}{ADC(2)}{second-order algebraic diagrammatic construction}
\newacronym{mp2}{MP2}{second-order M{\o}ller-Plesset perturbation theory}
\newacronym{mrci}{MRCI}{multireference configuration interaction}
\newacronym{nevpt2}{NEVPT2}{$n$-electron valence state second-order perturbation theory}

\usepackage{physics, calrsfs, mathtools, bm}
\DeclareMathAlphabet{\pazocal}{OMS}{zplm}{m}{n}

\newcommand*{\zod}{{\check{\gamma}}}
\newcommand*{\srdm}{\gamma}
\newcommand*{\effrdm}{\breve{\gamma}}
\newcommand*{\sardm}{\zod}

\DeclareMathOperator{\hess}{\vb{H}}

\newcommand*{\ot}{E^{\mathrm{ot}}}
\newcommand*{\oth}{F}
\newcommand*{\xc}{E^{\mathrm{xc}}}
\newcommand*{\otk}{\epsilon^{\mathrm{ot}}}
\newcommand*{\xck}{\epsilon^{\mathrm{xc}}}
\newcommand*{\fot}{\vb{f}^\mathrm{ot}}
\newcommand*{\fxc}{\vb{f}^\mathrm{xc}}
\newcommand*{\vot}{\vb{v}^\mathrm{ot}}
\newcommand*{\vxc}{\vb{v}^\mathrm{xc}}

\newcommand*{\rhop}{\rho^\prime}
\newcommand*{\Pip}{\Pi^\prime}
\newcommand*{\vrho}{\vec{\rho}}

\newcommand*{\vtrho}{\vec{\tilde{\rho}}}
\newcommand*{\rhoa}{\rho_\uparrow}
\newcommand*{\rhob}{\rho_\downarrow}
\newcommand*{\rhopa}{\rho^\prime_\uparrow}
\newcommand*{\rhopb}{\rho^\prime_\downarrow}
\newcommand*{\sigmaaa}{\sigma_{\uparrow\uparrow}}
\newcommand*{\sigmaab}{\sigma_{\uparrow\downarrow}}
\newcommand*{\sigmabb}{\sigma_{\downarrow\downarrow}}
\newcommand*{\tzeta}{\zeta_\mathrm{t}}
\newcommand*{\ftzeta}{\zeta_\mathrm{ft}}
\newcommand*{\ftzetap}{\zeta_\mathrm{ft}^\prime}

\newcommand*{\transpose}[1]{\ensuremath{#1^{\top}}}
\DeclareMathOperator{\jacobian}{\vb{J}}

\newcommand*{\PySCF}{\textsc{PySCF}\xspace}
\newcommand*{\PySCFforge}{\textsc{PySCF-forge}\xspace}
\newcommand*{\OpenMolcas}{\textsc{OpenMolcas}\xspace}
\newcommand*{\mrh}{\textsc{mrh}\xspace}
\newcommand*{\geomeTRIC}{\textsc{geomeTRIC}\xspace}
\newcommand*{\libxc}{\textsc{libxc}\xspace}
\newcommand*{\libcint}{\textsc{libcint}\xspace}

\begin{document}

\title{Analytic Nuclear Gradients for Complete Active Space Linearized Pair-Density Functional Theory}

\author{Matthew R. Hennefarth} 
\author{Matthew R. Hermes} \affiliation{Department of Chemistry and Chicago
	Center for Theoretical Chemistry, University of Chicago, Chicago, IL 60637,
	USA}

\author{Donald G. Truhlar} \email[corresponding author: ]{truhlar@umn.edu}
\affiliation{Department of Chemistry, Chemical Theory Center, and Minnesota
	Supercomputing Institute, University of Minnesota, Minneapolis, MN 55455-0431,
	USA}

\author{Laura Gagliardi} \email[corresponding author:
]{lgagliardi@uchicago.edu} \affiliation{Department of Chemistry, Pritzker
	School of Molecular Engineering, The James Franck Institute, and Chicago Center
	for Theoretical Chemistry, University of Chicago, Chicago, IL 60637, USA}
\affiliation{Argonne National Laboratory, 9700 S. Cass Avenue, Lemont, IL
	60439, USA}

\date{March 14, 2024}

\begin{abstract}
	Accurately modeling photochemical reactions is difficult due to the presence of conical intersections and locally avoided crossings as well as the inherently multiconfigurational character of excited states. As such, one needs a multi-state method that incorporates state interaction in order to accurately model the potential energy surface at all nuclear coordinates. The recently developed linearized pair-density functional theory (L-PDFT) is a multi-state extension of multiconfiguration PDFT, and it has been shown to be a cost-effective post-MCSCF method (as compared to more traditional and expensive multireference many-body perturbation methods or multireference configuration interaction methods) that can accurately model potential energy surfaces in regions of strong nuclear-electronic coupling in addition to accurately predicting Franck-Condon vertical excitations. In this paper, we report the derivation of analytic gradients for L-PDFT and their implementation in the \PySCFforge software, and we illustrate the utility of these gradients for predicting ground- and excited-state equilibrium geometries and adiabatic excitation energies for formaldehyde, \textit{s-trans}-butadiene, phenol, and cytosine.
\end{abstract}

\maketitle

\section{Introduction}

Accurate characterization and modeling of electronically excited states is important for a variety of chemical and biochemical processes including light-harvesting \cite{GraetzelSolar2005, ZhugayevychTheoretical2015, ProppeBioinspiration2020, CroceLight2020, McCuskerElectronic2019}, photochemistry \cite{DanielPhotochemistry2015, MaiMolecular2020}, photocatalysis \cite{HerrmannPhotocatalysis2017, RichardsPhoton2021}, vision \cite{WandShedding2013, KandoriRetinal2020}, and UV damage to DNA \cite{SobolewskiExcited2002, PlasserElectronic2014, ImprotaQuantum2016}. However, since excited states are typically strongly multiconfigurational, a multireference method is necessary for quantitative and qualitative accuracy. Additionally, since conical intersections and locally avoided crossings are common topological features of excited-state potential energy surfaces, the method should also be able to properly incorporate state interaction so that states of the same symmetry do not unphysically cross or cross on surfaces of the wrong dimensionality.

While state-averaged complete active space self-consistent field (CASSCF) theory \cite{RoosComplete1980, RoosComplete1987} is a multireference method that can properly account for static correlation, it does not include dynamic correlation outside of the active space. Multireference many-body perturbation theories such as CAS second-order perturbation theory (CASPT2) \cite{AnderssonSecond1990} or \gls{nevpt2} \cite{AngeliIntroduction2001} are able to recover external dynamic correlation; however, they are computationally very expensive. Multiconfiguration pair-density functional theory (MC-PDFT) \cite{LiManniMulticonfiguration2014, GhoshCombining2018, ZhouElectronic2022} is an alternative post-MCSCF approach that combines wave function theory and density functional theory where the final electronic energy of a multiconfigurational wave function is computed using an on-top energy functional for the nonclassical component of the energy. MC-PDFT recovers the missing correlation energy with an accuracy similar to CASPT2 \cite{HoyerMulticonfiguration2016} and \gls{nevpt2} \cite{KingLarge2022} but with a significantly reduced computational cost.

MC-PDFT is inaccurate in regions of strong nuclear-electronic coupling because it is a single-state method \cite{SandState2018, BaoMulti2020, BaoCompressed2020}. Linearized PDFT (L-PDFT) \cite{HennefarthLinearized2023} is a recently developed multi-state extension of MC-PDFT that incorporates state interaction by defining an effective Hamiltonian that is a functional of a set of densities. It has been shown to be as accurate as extended multi-state CASPT2 (XMS-CASPT2) \cite{GranovskyExtended2011} in modeling potential energy surfaces near conical intersections and locally avoided crossings \cite{HennefarthLinearized2023}, and it is as accurate as \gls{nevpt2} for predicting Franck-Condon vertical excitations \cite{HennefarthLinearized2023a}. L-PDFT also has the benefit of being computationally faster than MC-PDFT since its computational cost is independent of the number of states included in the model space, making it an excellent method to study photochemical reactions and dynamics.

Nuclear gradients are used to optimize molecular geometries to study vertical and adiabatic excitations and to perform molecular dynamics simulations. While it is always possible to calculate gradients numerically, it is often very slow to converge the gradients with respect to the stepsize. Here we report the derivation and implementation of analytic gradients for L-PDFT based on a SA-CASSCF reference wave function. Since the L-PDFT energy is not fully variational with respect to all wave function parameters, we use a Lagrangian-based approach similar to SA-CASSCF gradients \cite{StaalringAnalytical2001}, SA-MC-PDFT gradients \cite{SandAnalytic2018}, state-specific MC-PDFT (SS-MC-PDFT) gradients \cite{ScottAnalytic2020}, and MC-PDFT gradients with density fitting \cite{ScottAnalytic2021}. We then validate our implementation by comparing the analytic gradients to numerical gradients for the diatomic systems \ce{HeH+} and \ce{LiH}. Finally, we show the utility of L-PDFT gradients in predicting both ground- and excited-state geometries, as well as vertical and adiabatic excitation energies for formaldehyde, phenol, \textit{s-trans}-butadiene, and cytosine.

\section{Theory}

Throughout this manuscript, lowercase roman letters $p,q,r,s,t,u,v,w,x,y$ indicate general spatial \glspl{mo}, and lowercase Greek letters $\tau, \mu,\nu,\xi$ indicate \gls{ao} basis functions. $I,J$ refer to CASSCF eigenstates in the state-averaged space (which is taken to be the same as the model space); $\Lambda, \Gamma$ label L-PDFT eigenstates (within the model space); and $M,N$ label CASSCF eigenstates within the complementary part of the state-averaged space (outside the model space). $\kappa$ is used for \gls{mo} rotations, $\lambda$ for general nuclear coordinates, and $P$ for the state-transfer operator. Boldfaced variables are tensors (vectors, matrices, etc.). Einstein summation notation is used throughout (repeated indices are summed implicitly). An efficient implementation of the following equations should make use of the partitioning of orbitals into inactive, active, and virtual, and our code does this; however, our derivation presented in this manuscript does not account for such partitioning for simplicity.

\subsection{L-PDFT}

The L-PDFT energy \cite{HennefarthLinearized2023} of a given state $\ket{\Gamma}$ is defined as the first-order Taylor expansion of the MC-PDFT energy expression \cite{LiManniMulticonfiguration2014} in the one- and two-\gls{rdm} elements around the zero-order density ($\zod$).
\begin{equation}
	E^\mathrm{L-PDFT}_\Gamma = E^\mathrm{PDFT}\bqty{\zod} + \eval{\pdv{E^\mathrm{PDFT}}{\gamma^p_q}}_{\zod}\Delta^p_q + \eval{\pdv{E^\mathrm{PDFT}}{\gamma^{pr}_{qs}}}_{\zod}\Delta^{pr}_{qs}
\end{equation}
\begin{subequations}
	\begin{align}
		\Delta^p_q       & = \srdm^p_q - \zod^p_q             \\
		\Delta^{pr}_{qs} & = \srdm^{pr}_{qs} - \zod^{pr}_{qs}
	\end{align}
\end{subequations}
Here, $\gamma^p_q$ and $\gamma^{pr}_{qs}$ are the one- and two-\gls{rdm} elements of state $\ket{\Gamma}$, $\zod^p_q$ and $\zod^{pr}_{qs}$ are the one- and two-\gls{rdm} elements of the zero-order density, $\Delta$ represents the difference in \gls{rdm} elements between the state $\ket{\Gamma}$ and zero-order density, and $E^\mathrm{PDFT}\bqty{\zod}$ is the MC-PDFT energy expression evaluated with the zero-order \gls{rdm} elements.
\begin{equation}
	\label{eq:mcpdft-energy}
	E^\mathrm{PDFT}\bqty{\zod} = h^q_p\zod^p_q + \frac{1}{2}\pazocal{J}^q_p\bqty{\zod}\zod^p_q + \ot\bqty{\vrho_\zod}+V^\mathrm{nuc}
\end{equation}
\begin{equation}
	\pazocal{J}^q_p\bqty{\zod} = g^{qs}_{pr}\zod^r_s
\end{equation}
$h^q_p$ and $g^{qs}_{pr}$ are the normal one- and two-electron integrals respectively, $V^\mathrm{nuc}$ is the nuclear-nuclear repulsion, $\pazocal{J}^q_p\bqty{\zod}$ is the Coulomb interaction of the density $\zod$, and $\ot\bqty{\vrho_{\zod}}$ is an on-top energy functional that depends on the collective density variables $\vrho_\zod$. 
\begin{equation}
	\transpose{\vrho_\zod} = \begin{bmatrix}
		\rho_\zod & \Pi_\zod & \rhop_\zod & \Pip_\zod
	\end{bmatrix}
\end{equation}
The density ($\rho$), on-top pair density ($\Pi$), and their gradients are generated through the one- and two-\gls{rdm} elements as
\begin{equation}
	\rho_\zod = \phi_p\zod^p_q\phi^q
\end{equation}
\begin{equation}
	\Pi_\zod = \frac{1}{2}\phi_p\phi_r\zod^{pr}_{qs}\phi^q\phi^s
\end{equation}
\begin{equation}
	\rhop_\zod = \pqty{\phi^\prime_p\phi^q + \phi_p\phi^{\prime q}}\zod^p_q
\end{equation}
\begin{equation}
	\begin{split}
		\Pip_\zod = \frac{1}{2}(&\phi^\prime_p\phi_r\phi^q\phi^s + \phi_p\phi^\prime_r\phi^q\phi^s \\
		&+ \phi_p\phi_r\phi^{\prime q}\phi^s + \phi_p\phi_r\phi^q\phi^{\prime s} )\zod^{pr}_{qs}
	\end{split}
\end{equation}
Note that $\rho$, $\Pi$, and $\phi_p$ are all functions of one three-dimensional variable $\vb{r}$; and $\phi_p$ is the $p$'th \gls{mo}. \Cref{appendex:translated-functionals} describes how the current generation of on-top functionals are evaluated using existing \gls{ks} functionals via translated and fully-translated schemes.

The L-PDFT energy is expressible as \cite{HennefarthLinearized2023}
\begin{equation}
	\begin{split}
		E^\mathrm{L-PDFT}_{\Gamma} =& h^q_p\srdm^p_q + \pqty{\pazocal{J}^q_p\bqty{\zod} + V^q_p\bqty{\vrho_\zod}}\Delta^p_q + \frac{1}{2}v^{qs}_{pr}\bqty{\vrho_\zod}\Delta^{pr}_{qs} \\
		&+ \ot\bqty{\vrho_{\zod}} + \frac{1}{2}\pazocal{J}^q_p\bqty{\zod}\zod^p_q + V^\mathrm{nuc}
	\end{split}
\end{equation}
\begin{subequations}
	\begin{align}
		V^q_p       & = V^q_p\bqty{\vrho_\zod} =\eval{\pdv{\ot}{\gamma^p_q}}_{\vrho_\zod}               \\
		v^{qs}_{pr} & = v^{qs}_{pr}\bqty{\vrho_\zod} = 2\eval{\pdv{\ot}{\gamma^{pr}_{qs}}}_{\vrho_\zod}
	\end{align}
\end{subequations}
The L-PDFT energy contains derivatives of the on-top functional with respect to the one- ($V^p_q$) and two-\gls{rdm} ($v^{pr}_{qs}$) elements (which we call the one- and two-electron on-top potential terms \cite{SandAnalytic2018, ScottAnalytic2020}).

In L-PDFT, all functional terms ($\ot, V^q_p, v^{qs}_{pr}$) are evaluated only at the zero-order density. In practice, the zero-order density is taken to be the weighted average of densities within the state-averaged manifold.
\begin{subequations}
	\begin{align}
		\zod^p_q       & = \omega_I\expval{\hat{E}^p_q}{I}       \\
		\zod^{pr}_{qs} & = \omega_I\expval{\hat{e}^{pr}_{qs}}{I}
	\end{align}
\end{subequations}
$\hat{E}^p_q$ and $\hat{e}^{pr}_{qs}$ the one- and two-electron excitation operators respectively. We take $\omega_I$ to be the same weight as in the underlying SA-CASSCF or state-averaged complete active space configuration interaction calculation. For the analytic L-PDFT gradients, we require equal weights ($\omega_I = \omega_J$) so that the zero-order density is invariant to rotation among states within the model space, as is done in SA-CASSCF \cite{StaalringAnalytical2001}. We do not expect the limitation of L-PDFT analytical gradients to equal weights to impact the applicability of the method because almost all implemented SA-CASSCF analytical gradients also require equal weights in order to exploit the energy invariance to rotations within the state-averaged space.

Because the L-PDFT energy depends linearly on the one- and two-\gls{rdm} elements of the state, it is possible to express it as the expectation value of a Hermitian operator which we call the L-PDFT Hamiltonian ($\hat{H}^\mathrm{L-PDFT}$).
\begin{equation}
	E^\mathrm{L-PDFT}_{\Gamma} = \expval{\hat{H}^\mathrm{L-PDFT}}{\Gamma}
\end{equation}
\begin{equation}
	\hat{H}^\mathrm{L-PDFT} = \pqty{h^q_p + \pazocal{J}^q_p\bqty{\zod} + V^q_p}\hat{E}^p_q + \frac{1}{2}v^{qs}_{pr}\hat{e}^{pr}_{qs} + h^\mathrm{const}
\end{equation}
Here $h^\mathrm{const}$ is a constant term that only depends on $\zod$.
\begin{equation}
	h^\mathrm{const} = V^\mathrm{nuc} + \ot - \pqty{\frac{1}{2}\pazocal{J}^q_p\bqty{\zod} + V^q_p}\zod^p_q - \frac{1}{2}v^{qs}_{pr}\zod^{pr}_{qs}
\end{equation}

The final L-PDFT energies and states are the solutions to the eigenvalue equation of the operator  $\hat{H}^\mathrm{L-PDFT}$ projected within the model space spanned by the eigenvectors of the underlying SA-CASSCF calculation.
\begin{equation}
    \label{eq:lpdft-eigenvalue}
	\ket{\Gamma}\mel{\Gamma}{\hat{H}^\mathrm{L-PDFT}}{\Lambda} = \delta^\Lambda_{\Gamma}E^\mathrm{L-PDFT}_\Lambda \ket{\Gamma}
\end{equation}
Note that $\ket{\Gamma}$ and $\ket{I}$ span the same model space, but the state $\ket{I}$ are chosen such that they diagonalize the normal electronic Hamiltonian projected within that space.

\subsection{The L-PDFT Energy Lagrangian}

The Hellmann-Feynman theorem \cite{HellmannZur1933, HellmannEinfuhrung1937, FeynmanForces1939} implies that if an energy $E$ is stationary with respect to all parameters defining the wave function ($\ket{\Psi}$), then
\begin{equation}
	\dv{E}{\lambda} = \expval{\dv{\hat{H}}{\lambda}}{\Psi}
\end{equation}
As such, one does not have to account for the response of the wave function to a change in nuclear coordinate ($\lambda$).

The L-PDFT energy is not stationary with respect to \gls{mo} rotations and state rotations out of the model space; hence, the Hellmann-Feynman theory cannot be applied. However, one can avoid calculating the response of the wave function with respect to a nuclear displacement by using Lagrange’s method of undetermined multipliers. This requires us to enumerate the variables defining the wave function and the systems of equations that set them to their particular values.

Similarly to SA-CASSCF, the final L-PDFT eigenstates can be parameterized as
\begin{equation}
	\ket{\Gamma} = e^{\hat{P}^\Gamma}e^{\hat{\kappa}}\ket{0}
\end{equation}
where $\hat{\kappa}$ is the orbital rotation operator
\begin{equation}
	\hat{\kappa} = \sum_{p<q}\kappa^q_p\bqty{\hat{\vb{E}}-\hat{\vb{E}}^\dag}^p_q
\end{equation}
and $\hat{P}^\Gamma$ is the state transfer operator for state $\ket{\Gamma}$. $\hat{P}^\Gamma$ can be decomposed into rotations within the model space ($\hat{P}^\Gamma_\parallel$) and those outside the model space ($\hat{P}^\Gamma_\perp$).
\begin{equation}
	\hat{P}^\Gamma = \hat{P}_\perp^\Gamma + \hat{P}_\parallel^\Gamma
\end{equation}
\begin{equation}
	\hat{P}^\Gamma_\perp = P^\Gamma_M\pqty{\ketbra{M}{\Gamma} - \ketbra{\Gamma}{M}}
\end{equation}
\begin{equation}
	\hat{P}^\Gamma_\parallel = P^\Gamma_\Lambda\pqty{\ketbra{\Lambda}{\Gamma} - \ketbra{\Gamma}{\Lambda}}
\end{equation}
Since our reference wave function comes from a SA-CASSCF calculation, the parameters $\kappa^q_p$ and $P^\Gamma_M$ are optimized with respect to the SA-CASSCF energy ($E^\mathrm{SA}$).
\begin{equation}
	\label{eq:cas-mo-optimization}
	\pdv{E^\mathrm{SA}}{\kappa^q_p} = 0
\end{equation}
\begin{equation}
	\label{eq:cas-model-space-optimization}
	\pdv{E^\mathrm{SA}}{P^\Gamma_M} = 0
\end{equation}
The $P^\Gamma_\Lambda$ parameters are determined by diagonalizing $\hat{H}^\mathrm{L-{PDFT}}$, which is equivalent to making the energies stationary with respect to interstate rotations.
\begin{equation}
	\label{eq:lpdft-stationary-condition}
	\pdv{E^\mathrm{L-PDFT}_\Gamma}{P^\Gamma_\Lambda} = 0
\end{equation}
For equal weights, $E^\mathrm{SA}$ is invariant to rotation within the model space, and hence
\begin{equation}
	\label{eq:cas-invariant-model-space-rotations}
	\dv{}{x}\pqty{\pdv{E^\mathrm{SA}}{P_\Lambda^\Gamma}} = 0
\end{equation}
for any parameter $x$ \cite{StaalringAnalytical2001}.

Our constraints are given by \cref{eq:cas-mo-optimization,eq:cas-model-space-optimization}; therefore, our Lagrangian for state $\ket{\Gamma}$ takes the form
\begin{equation}
	\label{eq:lpdft-lagrangian}
	\mathcal{L}_\Gamma = E_\Gamma^\mathrm{L-PDFT} + \bar{\bm{\kappa}}\cdot\nabla_{\bm{\kappa}}E^\mathrm{SA}+\bar{\vb{P}}^\perp\cdot\nabla_{\vb{P}^\perp}E^\mathrm{SA}
\end{equation}
As a reminder $\bar{\bm{\kappa}}$ and $\bar{\vb{P}}^\perp$ are the associated Lagrange multipliers for the orbital rotations and state rotations out of the model space respectively. Both $\bar{\bm{\kappa}}$ and $\bar{\vb{P}}^\perp$ are determined by making the Lagrangian stationary with respect to $\kappa^q_p$ and $P^\Lambda_M$.
\begin{equation}
	\label{eq:lagrange-stationary-condition}
	\pdv{\mathcal{L}_\Gamma}{\kappa^q_p} = 0 = \pdv{\mathcal{L}_\Gamma}{P^\Lambda_M}
\end{equation}
As indicated by \cref{eq:lpdft-stationary-condition,eq:cas-invariant-model-space-rotations}, $\mathcal{L}_\Gamma$ is already invariant to rotations within the model space, and therefore does not need to be accounted for. Substituting \cref{eq:lpdft-lagrangian} into \cref{eq:lagrange-stationary-condition} yields a system of coupled linear equations.
\begin{equation}
	\label{eq:lagrange-mult-matrix}
	\begin{bmatrix}
		\nabla_{\bm{\kappa}}E^\mathrm{L-PDFT}_\Gamma \\
		\nabla_{\vb{P}^\perp}E_{\Gamma}^\mathrm{L-PDFT}
	\end{bmatrix} =-\begin{bmatrix}
		\hess_{\bm{\kappa} \bm{\kappa}}^{E^\mathrm{SA}}  & \hess_{\bm{\kappa} \vb{P}^\perp}^{E^\mathrm{SA}}   \\
		\hess_{\vb{P}^\perp \bm{\kappa}}^{E^\mathrm{SA}} & \hess_{\vb{P}^\perp  \vb{P}^\perp}^{E^\mathrm{SA}}
	\end{bmatrix}\begin{bmatrix}
		\bar{\bm{\kappa}} \\
		\bar{\vb{P}}^\perp
	\end{bmatrix}
\end{equation}
The left-hand side is the energy response of the L-PDFT energy, and the first factor on the right is the Hessian of $E^\mathrm{SA}$ with respect to $\kappa^q_p$ and $P^\Lambda_M$; these terms will be discussed in \cref{sec:lpdft-energy-response}. \Cref{eq:lpdft-lagrangian,eq:lagrange-mult-matrix} are almost identical to those presented in the SA-CASSCF analytic gradients \cite{StaalringAnalytical2001}, with the difference being that the wave function energy response terms on the left side have been replaced with the L-PDFT energy response terms. As such, \cref{eq:lagrange-mult-matrix} can be solved using the standard SA-CASSCF preconditioned conjugate gradient iterative solver \cite{PressNumerical1992, BernhardssonDirect1999}.

\subsection{Energy Response \label{sec:lpdft-energy-response}}

The SA-CASSCF Hessian that appears in \cref{eq:lagrange-mult-matrix} is well established \cite{HelgakerMolecular2000, StaalringAnalytical2001} and unchanged in this derivation. In practice, we evaluate $\hess^{E^\mathrm{SA}}$ within the L-PDFT eigenstate basis, rather than the SA-CASSCF eigenstate basis, which slightly modifies the form of the equation as is described in \cref{appendex:modified-casscf-hessian}.

As seen on the left side of \cref{eq:lagrange-mult-matrix}, we need the response of the L-PDFT energy with respect to orbital and CI rotations. However, it is important to realize that $E^\mathrm{L-PDFT}_{\Gamma}$ depends on the model space, and changing either the orbitals or CI parameters for any state within the model space changes the zero-order density. Hence, we break each component into an explicit and implicit part; the explicit part represents the response of only the state, whereas the implicit part accounts for changes due to the zero-order density.

\subsubsection{Explicit Dependence}

The derivatives of $\srdm$ with respect to $\kappa^q_p$ are given by
\begin{subequations}
	\label{eq:density-response-mo}
	\begin{equation}
		\pdv{\srdm^{p}_q}{\kappa^y_x} = \pqty{\delta^x_{q}\srdm^p_{y} + \delta^p_{x}\srdm^y_{q}} - \pqty{\delta^y_{q}\srdm^p_{x} + \delta^p_{y}\srdm^x_{q}} 
	\end{equation}
	\begin{equation}
		\begin{split}
			\pdv{\srdm^{pr}_{qs}}{\kappa^y_x} =& \pqty{\delta_x^p\srdm_{qs}^{yr} +\delta^x_q\srdm_{ys}^{pr} +\delta^r_x\srdm_{qs}^{py} + \delta_s^x\srdm_{qy}^{pr}}\\
			&- \pqty{\delta_y^{p}\srdm_{qs}^{xr} + \delta_q^{y}\srdm_{xs}^{pr} +\delta_y^{r}\srdm_{qs}^{px} + \delta_s^{y}\srdm_{qx}^{pr}}
		\end{split}
	\end{equation}
\end{subequations}
Therefore,
\begin{equation}
	\pdv{E^\mathrm{L-PDFT}_\Gamma}{\bm{\srdm}}\cdot\pdv{\bm{\srdm}}{\kappa^y_x} = 2\bqty{\mathcal{F}_\mathrm{ex} - \mathcal{F}^{\dag}_{\mathrm{ex}}}^x_y
\end{equation}
where $\mathcal{F}_\mathrm{ex}$ is the explicit part of the L-PDFT generalized Fock matrix.
\begin{equation}
	\bqty{\mathcal{F}_\mathrm{ex}}^x_y = \pqty{h^q_y+\pazocal{J}^q_y\bqty{\zod}+V^q_y}\srdm_{q}^{x} + v^{qs}_{yr}\srdm^{xr}_{qs}
\end{equation}

The derivative of $\srdm$ with respect to $P^\Lambda_M$ is given by
\begin{subequations}
	\label{eq:rdm-response-to-model-space-rotation}
	\begin{align}
		\pdv{\srdm^{p}_{q}}{P^\Lambda_M}   & = 2\delta^\Gamma_\Lambda \srdm^{M p}_{\Lambda q}   \\
		\pdv{\srdm^{pr}_{qs}}{P^\Lambda_M} & = 2\delta^\Gamma_\Lambda \srdm^{M pr}_{\Lambda qs}
	\end{align}
\end{subequations}
where $\gamma^{Mp}_{\Lambda q}$ and $\gamma^{Mpr}_{\Lambda qs}$ are the transition density matrix elements from $\ket{M}$ to $\ket{\Lambda}$.
\begin{subequations}
	\begin{align}
		\gamma^{Mp}_{\Lambda q}   & = \mel{\Lambda}{\hat{E}^p_q}{M}       \\
		\gamma^{Mpr}_{\Lambda qs} & = \mel{\Lambda}{\hat{e}^{pr}_{qs}}{M}
	\end{align}
\end{subequations}
Therefore, the explicit response of the L-PDFT energy to state rotations out of the model space is given by
\begin{equation}
	\pdv{E^\mathrm{L-PDFT}_\Gamma}{\bm{\srdm}}\cdot \pdv{\bm{\srdm}}{P^\Lambda_M} = 2\delta_\Lambda^\Gamma\mel{\Lambda}{\hat{H}^\mathrm{L-PDFT}}{M}
\end{equation}
In general, these expressions are similar to the SA-CASSCF response equations \cite{StaalringAnalytical2001} but with modified one- and two-electron integrals.

\subsubsection{Implicit Dependence}

Since the L-PDFT energy directly depends on the first derivatives of $\ot$ with respect to $\srdm$, the energy response involves explicit second derivatives. We define the elements of the Hessian of $\ot$ with respect to the \gls{rdm} elements as
\begin{subequations}
	\label{eq:on-top-hessian-def}
	\begin{align}
		\oth_{p,r}^{q,s}                        & = \eval{\pdv[2]{\ot}{\gamma_{q}^{p}}{\gamma_s^r}}_{\vrho_\zod}          \\
		\oth_{p,rt}^{q,su} = \oth_{rt,p}^{su,q} & = 2\eval{\pdv[2]{\ot}{\gamma_q^{p}}{\gamma_{su}^{rt}}}_{\vrho_\zod}     \\
		\oth_{pr,tv}^{qs,uw}                    & = 4\eval{\pdv[2]{\ot}{\gamma_{qs}^{pr}}{\gamma_{uw}^{tv}}}_{\vrho_\zod}
	\end{align}
\end{subequations}
Since $\ot$ is the integral of the on-top kernel ($\otk$) over all space,
\begin{equation}
	\ot\bqty{\vrho} = \int \otk\bqty{\vrho\pqty{\vb{r}}} \dd\vb{r}
\end{equation}
we can move the derivatives of \cref{eq:on-top-hessian-def} inside the integral. Applying the chain rule twice gives
\begin{equation}
	\label{eq:ot-hessian}
	\begin{split}
		\vb{\oth} =& \begin{bmatrix}
			\Bqty{\oth^{q,u}_{p,t}}   & \Bqty{\oth^{qs,u}_{pr,t}}   \\
			\Bqty{\oth^{u,qs}_{t,pr}} & \Bqty{\oth^{qs,uw}_{pr,tv}}
		\end{bmatrix} \\
		=& \int \dd\vb{r}\begin{bmatrix}
			\Bqty{\pdv{\vrho}{\gamma^p_q}} \\ 2\Bqty{\pdv{\vrho}{\gamma^{pr}_{qs}}}
		\end{bmatrix} \cdot \fot \cdot \begin{bmatrix}
			\Bqty{\pdv{\vrho}{\gamma^t_u}} & 2\Bqty{\pdv{\vrho}{\gamma^{tv}_{uw}}}
		\end{bmatrix}
	\end{split}
\end{equation}
\begin{subequations}
	\begin{align}
		\pdv{\vrho}{\gamma^p_q}       & = \begin{bmatrix}
			                                  \phi_p\phi^q          \\
			                                  0                     \\
			                                  2 \phi_p^\prime\phi^q \\
			                                  0
		                                  \end{bmatrix}                                                             \\
		\pdv{\vrho}{\gamma^{pr}_{qs}} & = \frac{1}{2}\begin{bmatrix}
			                                             0 \\ \phi_p\phi_r\phi^q\phi^s \\ 0 \\ 4 \phi^\prime_p\phi_r\phi^q\phi^s
		                                             \end{bmatrix}
	\end{align}
\end{subequations}
where $\fot$ is the Hessian of the on-top potential kernel with elements
\begin{equation}
	\fot = \begin{bmatrix}
		\pdv[2]{\otk}{\rho}     & \pdv{\otk}{\rho}{\Pi}  & \pdv{\otk}{\rho}{\rhop} & \pdv{\otk}{\rho}{\Pip}  \\
		\pdv{\otk}{\rho}{\Pi}   & \pdv[2]{\otk}{\Pi}     & \pdv{\otk}{\Pi}{\rhop}  & \pdv{\otk}{\Pi}{\Pip}   \\
		\pdv{\otk}{\rho}{\rhop} & \pdv{\otk}{\Pi}{\rhop} & \pdv[2]{\otk}{{\rhop}}  & \pdv{\otk}{\rhop}{\Pip} \\
		\pdv{\otk}{\rho}{\Pip}  & \pdv{\otk}{\Pi}{\Pip}  & \pdv{\otk}{\rhop}{\Pip} & \pdv[2]{\otk}{{\Pip}}
	\end{bmatrix}
\end{equation}
See \cref{appendix:translated-hessian} for a description of how $\fot$ is evaluated.

As we will see, $\vb{F}$ is always contracted with the $\Delta$ density elements. We define the Hessian-vector product $\bm{\Delta}\cdot\vb{\oth}$ to be the on-top gradient response.
\begin{subequations}
	\begin{align}
		\bqty{\bm{\Delta}\cdot\vb{\oth}}^u_t       & = \oth^{q,u}_{p,t}\Delta^p_q + \frac{1}{2}\oth^{qs,u}_{pr,t}\Delta^{pr}_{qs}     \\
		\bqty{\bm{\Delta}\cdot\vb{\oth}}^{uw}_{tv} & = \oth_{p,tv}^{q,uw}\Delta^p_q + \frac{1}{2}\oth_{pr,tv}^{qs,uw}\Delta^{pr}_{qs}
	\end{align}
\end{subequations}
We directly compute the elements of $\bm{\Delta}\cdot\vb{\oth}$ on the grid by moving the contraction with $\Delta$ within the integral, thereby avoiding constructing tensors of up to rank 8. We see this by first noting that the density variables $\vrho$ are linear with respect to the \gls{rdm} elements
\begin{equation}
	\label{eq:density-variables-linear}
	\vrho_\Delta = \pdv{\vrho}{\gamma^p_q}\Delta^p_q + \pdv{\vrho}{\gamma^{pr}_{qs}}\Delta^{pr}_{qs} = \pdv{\vrho}{\bm{\gamma}}\cdot\bm{\Delta}
\end{equation}
Thus, contracting \cref{eq:ot-hessian} with the elements of the $\Delta$, we find that the elements of $\bm{\Delta}\cdot\vb{\oth}$ are given by
\begin{subequations}
	\label{eq:on-top-gradient-response}
	\begin{align}
		\bqty{\bm{\Delta}\cdot\vb{\oth}}^q_p       & = \int \dd\vb{r} \transpose{\vrho}_\Delta \cdot \fot \cdot  \pdv{\vrho}{\gamma^p_q}       \\
		\bqty{\bm{\Delta}\cdot\vb{\oth}}^{qs}_{pr} & = 2\int \dd\vb{r} \transpose{\vrho}_\Delta \cdot \fot\cdot  \pdv{\vrho}{\gamma^{pr}_{qs}}
	\end{align}
\end{subequations}

The derivative of the $E^\mathrm{L-PDFT}_\Gamma$ with respect to the zero-order density matrix elements is given by
\begin{subequations}
	\begin{align}
		\pdv{E^\mathrm{L-PDFT}_\Gamma}{\zod^t_u}       & = \pazocal{J}^u_t\bqty{\Delta} +\bqty{\bm{\Delta}\cdot\vb{\oth}}^u_t \\
		\pdv{E^\mathrm{L-PDFT}_\Gamma}{\zod^{tv}_{uw}} & = \frac{1}{2}\bqty{\bm{\Delta}\cdot\vb{\oth}}^{uw}_{tv}
	\end{align}
\end{subequations}
Using \cref{eq:density-response-mo}, we see that the L-PDFT energy implicit response to \gls{mo} rotations is
\begin{equation}
	\pdv{E_\Gamma^\mathrm{L-PDFT}}{\bm{\zod}} \cdot \pdv{\bm{\zod}}{\kappa^y_x} = 2\bqty{\mathcal{F}_\mathrm{impl} - \mathcal{F}_{\mathrm{impl}}^{\dag}}^x_y
\end{equation}
where $\mathcal{F}_\mathrm{impl}$ is the implicit contribution to the L-PDFT generalized Fock matrix.
\begin{equation}
	\bqty{\mathcal{F}_\mathrm{impl}}^x_y = \pqty{\pazocal{J}^q_y\bqty{\Delta} + \bqty{\bm{\Delta}\cdot\vb{\oth}}^q_y}\zod^x_q + \bqty{\bm{\Delta}\cdot\vb{\oth}}_{yr}^{qs}\zod_{qs}^{xr}
\end{equation}
Additionally, if we consider \cref{eq:rdm-response-to-model-space-rotation}, we find that
\begin{subequations}
	\begin{align}
		\pdv{\check{\gamma}^p_q}{P^\Lambda_M}       & = 2\omega_\Lambda \gamma^{M p}_{\Lambda q}   \\
		\pdv{\check{\gamma}^{pr}_{qs}}{P^\Lambda_M} & = 2\omega_\Lambda \gamma^{M pr}_{\Lambda qs}
	\end{align}
\end{subequations}
Therefore, we have that the implicit response of $E^\mathrm{L-PDFT}_\Gamma$ to state rotations out of the model space is given by
\begin{equation}
	\pdv{E^\mathrm{L-PDFT}_\Gamma}{\bm{\zod}}\cdot\pdv{\bm{\zod}}{P^\Lambda_M} = 2\omega_\Lambda G^M_\Lambda
\end{equation}
where we have defined $G^M_\Lambda$ as
\begin{equation}
	G^M_\Lambda = \mel{\Lambda}{\hat{G}}{M}
\end{equation}
\begin{equation}
	\hat{G} = \pqty{\pazocal{J}^q_p\bqty{\Delta} + \bqty{\bm{\Delta}\cdot\vb{\oth}}^q_p}\hat{E}^p_q + \frac{1}{2}\bqty{\bm{\Delta}\cdot\vb{\oth}}^{qs}_{pr}\hat{e}^{pr}_{qs}
\end{equation}

\subsubsection{Response Summary}

Given the above decompositions, we have the L-PDFT energy response to orbitals rotations is given by
\begin{equation}
	\pdv{E^\mathrm{L-PDFT}_\Gamma}{\kappa^y_x} = 2\bqty{\mathcal{F} - \mathcal{F}^\dag}^x_y
\end{equation}
where $\mathcal{F}$ is the full L-PDFT generalized Fock matrix.
\begin{equation}
	\label{eq:lpdft-generalized-fock}
	\begin{split}
		\mathcal{F}^x_y =& \bqty{\mathcal{F}_\mathrm{ex} + \mathcal{F}_\mathrm{impl}}^x_y\\
		=& \pqty{h^q_y+\pazocal{J}^q_y\bqty{\zod}+V^q_y}\srdm^x_q + v^{qs}_{yr}\srdm^{xr}_{qs} \\
		&+ \pqty{\pazocal{J}^q_y\bqty{\Delta} + \bqty{\bm{\Delta}\cdot\vb{\oth}}^q_y}\zod^x_q + \bqty{\bm{\Delta}\cdot\vb{\oth}}^{qs}_{yr}\zod^{xr}_{qs}
	\end{split}
\end{equation}
Additionally, we have that the L-PDFT energy response to state rotations out of the model space is given by
\begin{equation}
	\label{eq:lpdft-rotation-response}
	\pdv{E^\mathrm{L-PDFT}_\Gamma}{P^\Lambda_M} = 2\pqty{\delta_\Lambda^\Gamma\bqty{\hat{H}^\mathrm{L-PDFT}}_{\Lambda}^M + \omega_\Lambda G_\Lambda^M}
\end{equation}
Note that when there is only one state within the model space, $\zod \to \srdm$ and the L-PDFT energy is exactly the same as the SS-MC-PDFT energy. Correspondingly, one can see that all of the implicit terms would vanish from \cref{eq:lpdft-generalized-fock,eq:lpdft-rotation-response}, which would result in the same response equations as derived for the SS-MC-PDFT gradients \cite{SandAnalytic2018}.

\subsection{Derivative of the Lagrangian}

Having solved for the Lagrange multipliers in \cref{eq:lpdft-lagrangian} via \cref{eq:lagrange-mult-matrix}, the gradient of $\mathcal{L}_\Gamma$ yields the same gradient as fully differentiating $E^\mathrm{L-PDFT}_\Gamma$. Differentiating \cref{eq:lpdft-lagrangian} with respect to nuclear displacements yields
\begin{equation}
	\label{eq:lagrangian-nuclear-derivative}
	\begin{split}
		\dv{\mathcal{L}_\Gamma}{\lambda} =& \dv{E^\mathrm{L-PDFT}_\Gamma}{\lambda}\\
		=& \expval{\dv{\hat{H}^\mathrm{L-PDFT}}{\lambda}}{\Gamma} + \bar{\bm{\kappa}} \cdot \nabla_{\bm{\kappa}}\pqty{\omega_\Lambda \expval{\dv{\hat{H}^\mathrm{el}}{\lambda}}{\Lambda}} \\
		&+ \bar{\vb{P}}^\perp \cdot \nabla_{\vb{P}^\perp} \pqty{\omega_\Lambda\expval{\dv{\hat{H}^\mathrm{el}}{\lambda}}{\Lambda}}
	\end{split}
\end{equation}

Both the electronic Hamiltonian and the L-PDFT Hamiltonian in \cref{eq:lagrangian-nuclear-derivative} are expressed within second quantization, which requires that the \glspl{mo} always be orthonormal. A change in nuclear coordinates can cause the underlying \glspl{mo} to no longer be orthogonal. In particular, if we consider the following \gls{ao} to \gls{mo} transformation at a fixed geometry ($\lambda_0$),
\begin{equation}
	\ket{\phi^p;\lambda_0} = C^p_\mu\ket{\phi^\mu;\lambda_0}
\end{equation}
then for any slight nuclear displacement $\delta$, the overlap between the perturbed \glspl{mo} at the this perturbed geometry is no longer orthonormal.
\begin{equation}
	\braket{\phi_p;\lambda_0+\delta}{\phi^q;\lambda_0+\delta} \neq \delta^q_p
\end{equation}
To ensure that our \glspl{mo} are orthogonal regardless of the geometry, we introduce the \gls{omo} picture \cite{Helgakersecond‐quantization1984} (using the L\"{o}wdin orthonormalization \cite{Loewdinnon1950}),
\begin{equation}
    \label{eq:omo-definition}
	\ket{\psi^{\tilde{p}}; \lambda} = \bqty{\vb{S}^{-\frac{1}{2}}\pqty{\lambda}}^{\tilde{p}}_p C^p_\mu \ket{\phi{^\mu};\lambda}
\end{equation}
where $\vb{S}\pqty{\lambda}$ is the overlap matrix in the \gls{mo} basis
\begin{equation}
	S^p_q\pqty{\lambda} = \braket{\phi_q;\lambda}{\phi^p;\lambda}
\end{equation}
At the reference geometry ($\lambda=\lambda_0$), the \glspl{omo} are the same as the \glspl{mo} since $\vb{S}\pqty{\lambda_0}$ is the identity matrix. \Cref{eq:omo-definition} also highlights that the transformation from \glspl{ao} to \glspl{omo} also depends on the nuclear coordinates. The derivatives of the one- and two-electron integrals in the \gls{omo} basis are given by
\begin{equation}
	\dv{h^{\tilde{q}}_{\tilde{p}}}{\lambda} = h^q_{p,\lambda} - \frac{1}{2}\pqty{S^x_{p,\lambda}h^q_x + S^q_{x,\lambda}h^x_p}
\end{equation}
\begin{equation}
	\dv{g^{\tilde{q}\tilde{s}}_{\tilde{p}\tilde{r}}}{\lambda} = g^{qs}_{pr,\lambda} - \frac{1}{2}\pqty{S^x_{p,\lambda} g_{xr}^{qs} + S^x_{r,\lambda} g_{px}^{qs} + S^q_{x,\lambda} g_{pr}^{xs} + S^s_{x,\lambda} g_{pr}^{qx}}
\end{equation}
where $h^q_{p,\lambda}$ and $g^{qs}_{pr,\lambda}$ are the derivatives of the one- and two-electron integrals in the \gls{mo} basis, and $S^p_{q,\lambda}$ is the derivative of the overlap matrix element. The term involving the derivative of the overlap elements is often called the ``connection'' or ``renormalization'' contribution. The one- and two-electron derivative integrals and the overlap matrix elements are obtained in the \gls{ao} basis as
\begin{equation}
	h^\mu_{\tau,\lambda} = \mel{\phi_{\tau,\lambda}}{\hat{h}}{\phi^\mu} + \mel{\phi_{\tau}}{\hat{h}}{\phi^\mu_{,\lambda}} + \mel{\phi_\tau}{\pdv{\hat{h}}{\lambda}}{\phi^\mu}
\end{equation}
\begin{equation}
	\begin{split}
		g_{\tau\nu,\lambda}^{\mu\xi} =& \mel{\phi_{\tau,\lambda}\phi_\nu}{\hat{g}}{\phi^\mu\phi^\xi} + \mel{\phi_{\tau}\phi_{\nu,\lambda}}{\hat{g}}{\phi^\mu\phi^\xi}\\
		&+ \mel{\phi_{\tau}\phi_{\nu}}{\hat{g}}{\phi^\mu_{,\lambda}\phi^\xi} + \mel{\phi_{\tau}\phi_{\nu}}{\hat{g}}{\phi^\mu\phi^\xi_{,\lambda}}
	\end{split}
\end{equation}
\begin{equation}
	S^\mu_{\tau,\lambda} = \braket{\phi_{\tau,\lambda}}{\phi^\mu} + \braket{\phi_\tau}{\phi^\mu_{,\lambda}}
\end{equation}
where $\phi_{\tau,\lambda}$ represents the partial derivative of $\phi_\tau$ with respect to the nuclear coordinate.
\begin{equation}
	\phi_{\tau,\lambda} = \pdv{\phi_\tau}{\lambda}
\end{equation}

Within the L-PDFT Hamiltonian, there are no explicit two-electron integrals, but only Coulomb integrals. The Coulomb derivative integrals, generated from a density $\gamma$, are given by
\begin{equation}
	\pazocal{J}^q_{p,\lambda}\bqty{\gamma} = \pqty{\mel{\phi_{p,\lambda} \phi_r}{\hat{g}}{\phi^q\phi^s} + \mel{\phi_{p} \phi_r}{\hat{g}}{\phi^{q,\lambda}\phi^s} }\gamma^r_s
\end{equation}
The derivative of the Coulomb contribution can be re-written in terms of the Coulomb derivative integrals. For example,
\begin{equation}
	\begin{split}
		\pdv{\pazocal{J}^q_p\bqty{\zod}}{\lambda}\Delta^p_q &= g^{qs}_{pr,\lambda}\zod^r_s\Delta^p_q \\
		&= \pazocal{J}^q_{p,\lambda}\bqty{\zod}\Delta^p_q + \pazocal{J}^q_{p,\lambda}\bqty{\Delta}\zod^p_q
	\end{split}
\end{equation}

Since the on-top energy and on-top potential terms depend on the collective density variables (which are constructed from the \glspl{omo}), they will also give a renormalization contributions \cite{SandAnalytic2018,ScottAnalytic2020}.
\begin{equation}
	\dv{\ot}{\lambda} = \ot_{,\lambda} - S^y_{x,\lambda}\pqty{V^p_y\zod^x_p + v^{qs}_{yr}\zod^{xr}_{qs}}
\end{equation}
\begin{subequations}
	\begin{align}
		\dv{V^{\tilde{q}}_{\tilde{p}}}{\lambda}\Delta^{\tilde{p}}_{\tilde{q}}                                     & = V^q_{p,\lambda}\Delta^p_q - S^y_{x,\lambda}\pqty{V^p_y\Delta^x_p + \bqty{\Delta\cdot\vb{F}}^p_y\zod^x_p}                            \\
		\dv{v^{\tilde{q}\tilde{s}}_{\tilde{p}\tilde{r}}}{\lambda}\Delta^{\tilde{p}\tilde{r}}_{\tilde{q}\tilde{s}} & = v^{qs}_{pr,\lambda}\Delta^{pr}_{qs} - 2S^y_{x,\lambda}\pqty{v^{qs}_{yr}\Delta^{xr}_{qs} + \bqty{\Delta\cdot\vb{F}}^{qs}_{yr}\zod^{xr}_{qs}}
	\end{align}
	\label{eq:dVlambda}
\end{subequations}

The explicit derivatives of the on-top energy ($\ot_{,\lambda}$) and one- and two-electron on-top potentials ($V^q_{p,\lambda}$ and $v^{qs}_{pr,\lambda}$) can be evaluated in various forms since the tensor contractions can be performed in different orders. Here we present the derivatives as they are implemented in our \PySCFforge implementation. The nuclear derivative of the collective density variables can be written as
\begin{equation}
	\pdv{\transpose{\vrho_\srdm}}{\lambda} = \begin{bmatrix}
		\pdv{\rho_\srdm}{\lambda} & \pdv{\Pi_\srdm}{\lambda} & \pdv{\rhop_\srdm}{\lambda} & \pdv{\Pip_\srdm}{\lambda}
	\end{bmatrix} + \delta_{\lambda}(\vb{r})\nabla\vrho_{\srdm}\cdot\vec{\vb{n}}_{\lambda}
\end{equation}
\begin{equation}
	\pdv{\rho_\srdm}{\lambda} = 2\phi_{p}\srdm^p_q\phi^q_{,\lambda}
\end{equation}
\begin{equation}
	\pdv{\Pi_\srdm}{\lambda} = 2\phi_{p,\lambda}\phi_r\srdm^{pr}_{qs}\phi^q\phi^s
\end{equation}
\begin{equation}
	\pdv{\rhop_\srdm}{\lambda} = 2\pqty{\phi^\prime_{p,\lambda}\phi^q + \phi^\prime_p \phi^q_{,\lambda}}\srdm^p_q
\end{equation}
\begin{equation}
	\pdv{\Pip_\srdm}{\lambda} = 2\pqty{\phi^\prime_{p,\lambda}\phi_r\phi^q\phi^s + 3\phi^\prime_p\phi_{r,\lambda}\phi^q\phi^s}\srdm^{pr}_{qs}
\end{equation}
where $\delta_\lambda(\vb{r})$ is 1 if $\vb{r}$ is evaluated at a grid point associated with the atom of the coordinate $\lambda$ and is 0 otherwise, and $\vec{\vb{n}}_{\lambda}$ is the Cartesian unit vector for the coordinate direction $\lambda$. The explicit derivative of the on-top energy with respect to nuclear coordinates is given as
\begin{equation}
	\ot_{,\lambda} = \int \dd\vb{r}\pqty{\vot \cdot \pdv{\vrho_{\zod}}{\lambda}} + \Bqty{\otk}_{\vb{r}\in \mathcal{G}} \cdot \pdv{\vec{w}}{\lambda}
\end{equation}
where $\mathcal{G}$ is the set of all grid points, $\vec{w}$ is the corresponding set of quadrature weights, $\vot$ is the derivative of the on-top kernel with respect to the density variables,
\begin{equation}
	\vot = \nabla_{\vrho}\otk = \begin{bmatrix}
		\pdv{\otk}{\rho} & \pdv{\otk}{\Pi} & \pdv{\otk}{\rhop} & \pdv{\otk}{\Pip}
	\end{bmatrix}
\end{equation}
and
\begin{equation}
	\transpose{\Bqty{\otk}_{\vb{r}\in \mathcal{G}}} =\begin{bmatrix}\otk(\vb{r}_1) & \otk(\vb{r}_2) & \ldots & \otk(\vb{r}_n)\end{bmatrix}
\end{equation}
is the on-top kernel evaluated at every grid point (as a vector). See \cref{appendix:translated-gradient} for how $\vot$ is evaluated. The nuclear derivative of the on-top potential is given as
\begin{subequations}
	\begin{align}
		\begin{split}
			V^q_{p,\lambda} =& \int \dd \vb{r} \pqty{\pdv{\transpose{\vrho}}{\gamma^p_q} \cdot \fot \cdot \pdv{\vrho_{\zod}}{\lambda} + \vot \cdot \pdv{\vrho}{\gamma^p_q}{\lambda}} \\
			&+ \Bqty{\vot \cdot \pdv{\vrho}{\gamma^p_q}}_{\vb{r}\in \mathcal{G}} \cdot  \pdv{\vec{w}}{\lambda}
		\end{split} \\
		\begin{split}
			v^{qs}_{pr,\lambda} &= 2\int \dd \vb{r} \pqty{\pdv{\transpose{\vrho}}{\gamma^{pr}_{qs}} \cdot \fot \cdot \pdv{\vrho_{\zod}}{\lambda} + \vot \cdot \pdv{\vrho}{\gamma^{pr}_{qs}}{\lambda}} \\
			&+ 2\Bqty{\vot \cdot \pdv{\vec{\rho}}{\gamma^{pr}_{qs}}}_{\vb{r}\in \mathcal{G}} \cdot  \pdv{\vec{w}}{\lambda}
		\end{split}
	\end{align}
\end{subequations}
However, because the density variables are linear with respect to the \gls{rdm} elements (\cref{eq:density-variables-linear}), we can move the contraction with $\Delta$ in \cref{eq:dVlambda} within the integral (just like in \cref{eq:on-top-gradient-response}). Additionally, to reduce quadrature loops, we evaluate the on-top energy derivative at the same time such that the nuclear derivative due to all on-top terms is given as
\begin{equation}
	\begin{split}
		\ot_{,\lambda} + V^{q}_{p,\lambda}&\Delta^{p}_{q} + \frac{1}{2}v^{qs}_{pr,\lambda}\Delta^{pr}_{qs} \\
		=& \int\dd\vb{r}\pqty{\transpose{\vrho_\Delta}\cdot\fot\cdot\pdv{\vrho_{\check{\gamma}}}{\lambda} + \vot\cdot\pdv{\vrho_\srdm}{\lambda}} \\
		&+ \Bqty{\otk + \vot\cdot\vrho_\Delta}_{\vb{r}\in\mathcal{G}} \cdot \pdv{\vec{w}}{\lambda}
	\end{split}
\end{equation}

\begin{table*}
	\caption{\label{tab:systems-studied} Systems studied along with the symmetry, basis set, number of states ($N$), number of active electrons ($n_\mathrm{e}$), active space orbitals, and the on-top functional used.}
	\begin{tabular*}{\textwidth}{@{\extracolsep{\fill}} l c c c c c c}
		\toprule\toprule
		System       & Sym.\footnote{Point group symmetry} & Basis Set                                                                                                           & {$N$}                                              & $n_\mathrm{e}$ & Active Orbitals                                                                            & Functional                                                                                                                                                                                                      \\
		\midrule
		\ce{HeH+}    & $C_1$                                & cc-pVDZ \cite{DunningGaussian1989}                                                                                  & 2                                                  & 2              & $\sigma,\sigma^*$                                                                          & tPBE \cite{LiManniMulticonfiguration2014, PerdewGeneralized1996}                                                                                                                                                   \\
		&                                      &                                                                                                                     &                                                    &                &                                                                                            & ftSVWN3\footnote{ftSVWN3 is a fully-translated \cite{CarlsonMulticonfiguration2015} local-spin-density approximation with the exchange functional being the one called Slater exchange \cite{BlochBemerkung1929, DiracNote1930} in \libxc and the correlation functional being correlation functional number 3 of \citet{VoskoAccurate1980}.} \\
		\ce{LiH}     & $C_1$                                & aug-cc-pVTZ \cite{DunningGaussian1989, KendallElectron1992}                                                          & 2                                                  & 2              & $\sigma,\sigma^*$                                                                          & tPBE \cite{LiManniMulticonfiguration2014,PerdewGeneralized1996}                                                                                                                                                   \\
		&                                      &                                                                                                                     &                                                    &                &                                                                                            & ftPBE \cite{CarlsonMulticonfiguration2015,PerdewGeneralized1996}                                                                                                                                                \\
		formaldehyde & $C_1$                                & jun-cc-pVTZ \cite{FellerRole1996, SchuchardtBasis2007, DunningGaussian1989, KendallElectron1992, PapajakConvergent2010} & 2                                                  & 12             & full valence                                                                               & tPBE \cite{LiManniMulticonfiguration2014, PerdewGeneralized1996}                                                                                                                                                   \\
		\textit{s-trans}-butadiene    & $C_{2h}$                             & jul-cc-pVTZ \cite{FellerRole1996, SchuchardtBasis2007, DunningGaussian1989, KendallElectron1992, PapajakConvergent2010} & 2 (\textsuperscript{1}A\textsubscript{g})          & 4              & $2(\pi,\pi^*)$                                                                             & tPBE \cite{LiManniMulticonfiguration2014, PerdewGeneralized1996}                                                                                                                                                   \\
		phenol       & $C_1$                                & jul-cc-pVDZ \cite{FellerRole1996, SchuchardtBasis2007, DunningGaussian1989, KendallElectron1992, PapajakConvergent2010} & 3                                                  & 12             & $3(\pi,\pi^*)$, p$_z$ of O                                                                 & tPBE \cite{LiManniMulticonfiguration2014, PerdewGeneralized1996}                                                                                                                                                   \\
		&                                      &                                                                                                                     &                                                    &                & $\sigma_\mathrm{OH}$, $\sigma^*_\mathrm{OH}$, $\sigma_\mathrm{CO}$, $\sigma^*_\mathrm{CO}$                                                                                                                                                                                                                   \\
		cytosine     & $C_s$                                & jul-cc-pVTZ \cite{FellerRole1996, SchuchardtBasis2007, DunningGaussian1989, KendallElectron1992, PapajakConvergent2010} & 3 (\textsuperscript{1}A\textsuperscript{$\prime$}) & 14             & $5\pi$, $3\pi^*$, and 2 lone-pairs                                                         & tPBE \cite{LiManniMulticonfiguration2014, PerdewGeneralized1996}\\
		\bottomrule\bottomrule
	\end{tabular*}
\end{table*}
We can therefore write the full gradient of the L-PDFT energy for state $\ket{\Gamma}$ as
\begin{equation}
	\begin{split}
		\dv{E^\mathrm{L-PDFT}_\Gamma}{\lambda} =& \expval{\dv{\hat{H}^\mathrm{L-PDFT}}{\lambda}}{\Gamma} \\
		&+ h^q_{p,\lambda}\effrdm^p_q + \frac{1}{2}g^{qs}_{pr,\lambda}\effrdm^{pr}_{qs} - S^y_{x,\lambda}\breve{\mathcal{F}}^x_y
	\end{split}
\end{equation}
where $\effrdm$ are the effective \gls{rdm} elements, which contain the Lagrange multiplier terms as
\begin{subequations}
	\label{eq:effective-rdm}
	\begin{align}
		\breve{\gamma}^p_q = & \pqty{\sardm^x_q\bar{\kappa}^p_x - \sardm^p_x\bar{\kappa}^x_q} + \omega^\Lambda \bar{P}^\Lambda_M \bqty{\hat{\vb{E}} + \hat{\vb{E}}^\dag}^{Mp}_{\Lambda q} \\
		\begin{split}
			\breve{\gamma}^{pr}_{qs} =& \pqty{\sardm^{xr}_{qs}\bar{\kappa}^p_x -\sardm^{pr}_{xs}\bar{\kappa}^x_q + \sardm^{px}_{qs}\bar{\kappa}^r_x -\sardm^{pr}_{qx}\bar{\kappa}^x_s} \\
			&+ \omega^\Lambda \bar{P}^\Lambda_M\bqty{\hat{\vb{e}} + \hat{\vb{e}}^\dag}^{Mpr}_{\Lambda qs}
		\end{split}
	\end{align}
\end{subequations}
and $\breve{\mathcal{F}}$ is the effective Fock matrix,
\begin{equation}
	\label{eq:effective-generalized-fock}
	\breve{\mathcal{F}}^\mu_\tau = h^\nu_\tau\effrdm^\mu_\nu + g^{\nu \xi}_{\tau \eta}\effrdm^{\mu \eta}_{\nu \xi}
\end{equation}
The L-PDFT Hellmann-Feynman contribution is given as
\begin{equation}
	\begin{split}
		\expval{\dv{\hat{H}^\mathrm{L-PDFT}}{\lambda}}{\Gamma} =& h^q_{p,\lambda}\srdm^p_q + \pazocal{J}^q_{p,\lambda}\bqty{\zod}\Delta^p_q + \pazocal{J}^q_{p,\lambda}\bqty{\srdm}\zod^p_q \\
		&+ V^\mathrm{nuc}_{,\lambda} + \ot_{,\lambda} + V^q_{p,\lambda}\Delta^p_q + \frac{1}{2}v^{qs}_{pr,\lambda}\Delta^{pr}_{qs} \\
		& - S^y_{x,\lambda}\mathcal{F}^x_y
	\end{split}
\end{equation}
where $\mathcal{F}$ is the generalized L-PDFT Fock matrix (defined in \cref{eq:lpdft-generalized-fock}). \Cref{eq:effective-rdm,eq:effective-generalized-fock} are essentially the same as those that appear in other Lagrange-based analytic gradient approaches \cite{StaalringAnalytical2001, CelaniAnalytical2003, SandAnalytic2018, ScottAnalytic2020}.

\section{Computational Methods}

All calculations used \PySCF \cite{SunPySCF2018, SunRecent2020} (Version 2.3, commit \texttt{v1.1-8104-g6c1ea86eb}) compiled with the \libxc \cite{MarquesLibxc2012, LehtolaRecent2018} (Version 6.1.0) and \libcint \cite{SunLibcint2015} (Version 6.0.0) libraries, \mrh \cite{Hermesmrh2024} (commit SHA-1 \texttt{b3185fe}), and \PySCFforge \cite{PySCF2024} (commit SHA-1 \texttt{c503f41}). Geometry optimizations were performed with the \geomeTRIC \cite{WangGeometry2016} package (version 1.0) within \PySCF. All PDFT calculations used a numerical quadrature grid size of 6 (80/120 radial points and 770/974 angular points for atoms of periods 1/2 respectively).  All L-PDFT calculations used the model space spanned by the SA-CASSCF eigenvectors to construct $\hat{H}^\mathrm{L-PDFT}$ and equal weights ($\omega_I = \omega_J$). System-specific computational details including symmetry, basis set, number of states in the model space, active space, and on-top functional used are summarized in \cref{tab:systems-studied}.

Numerical gradients were computed using the central difference method. Because there is a dependence on the step size ($\delta$), we calculated numerical gradients with differing $\delta$ and extrapolated to the $\delta\to0$ limit. Specifically, a linear regression of the numerical gradient versus $\delta^2$ is performed for a subset of data where the correlation coefficient $R^2$ is greater than 0.9, and the $y$-intercept is taken to be the extrapolated numerical gradient. 

Both numerical and analytic gradients suffer from numerical error since they rely on a wave function that is only converged to finite precision. For all comparisons between analytic and numerical gradients, we use an energy convergence threshold of $10^{-12}$ \unit{\hartree} and an orbital and CI rotation gradient threshold of $10^{-6}$. For convenience, we will refer to the numerical gradient as the reference for the rest of this manuscript (as we have done previously \cite{ScottAnalytic2020, BaoAnalytic2022}). We also define the \gls{ue} (in units of \unit{\hartree\per\bohr}) and \gls{re} (unitless) as
\begin{equation}
	\text{\gls{ue}} = \abs{\mathrm{Analytic} - \mathrm{Numerical}}
\end{equation}
\begin{equation}
	\text{\gls{re}} = \abs{\frac{\mathrm{Analytic} - \mathrm{Numerical}}{\mathrm{Numerical}}}
\end{equation}

\section{Results and Discussion}

\subsection{Validation of Analytic Gradients Using Diatomic Molecules}

We first test our analytic gradient implementation with translated and fully-translated functionals at a variety of points on the potential energy curves of two diatomic systems: \ce{HeH+} \cite{GlocklerHelium1933, BishopTheoretical1979, PeyerimhoffHartreeFockRoothaan2004, GuestenAstrophysical2019, NovotnyQuantum2019} and \ce{LiH} \cite{FallonPotential1960, LiA11978, PardoPade1986, StwalleySpectroscopy1993, TungVery2011}. Analytic and numerical gradients were computed for interatomic distances in the range  \qtyrange{0.4}{4.0}{\angstrom} with a step size of \qty{0.1}{\angstrom}. As both the common log of \gls{ue} and \gls{re} occur in roughly standard distributions, we present the errors as histograms.

\begin{figure}
	\centering
	\includegraphics[width=0.95\columnwidth]{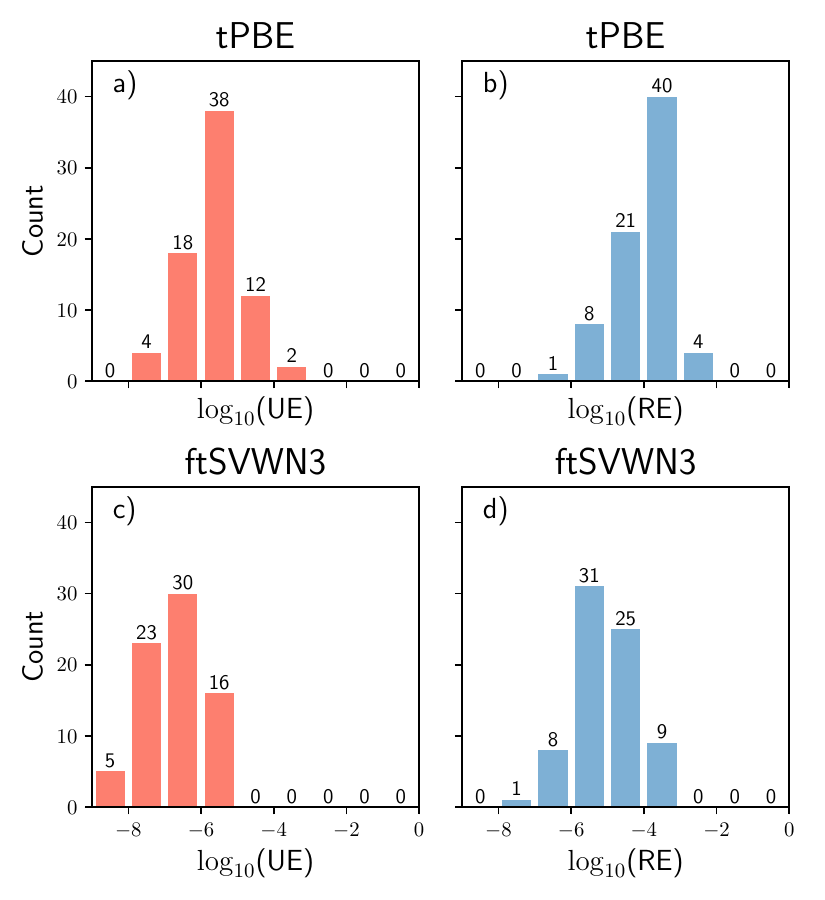}
	\caption{\label{fig:hhe-hist} Distribution of the common log of the unsigned (a,c) and relative (b,d) error of analytic gradients relative to numerical gradients for all states of \ce{HeH+} at various internuclear distances. The top row (a,b) is using the tPBE functional and bottom row (c,d) is using the ftSVWN3 functional.}
\end{figure}

\ce{HeH+} is the simplest possible system for which many terms in the programmable equations for L-PDFT analytic gradients do not vanish due to symmetry. The $\log_{10}$(\glsxtrshort{ue}) and $\log_{10}$(\glsxtrshort{re}) distributions for both the tPBE and ftSVWN3 functional are shown in \cref{fig:hhe-hist}. For the tPBE functional, the majority of the \glspl{ue} are below \qty{1e-5}{\hartree\per\bohr} whereas for the ftSVWN3 functional they are below \qty{1e-6}{\hartree\per\bohr}. The \gls{re} is fairly constant throughout the potential energy curve for both functionals (Fig. S1 and S2) with the \gls{re} for tPBE being an order of magnitude greater than for ftSVWN3.

\begin{figure}
	\centering
	\includegraphics[width=0.95\columnwidth]{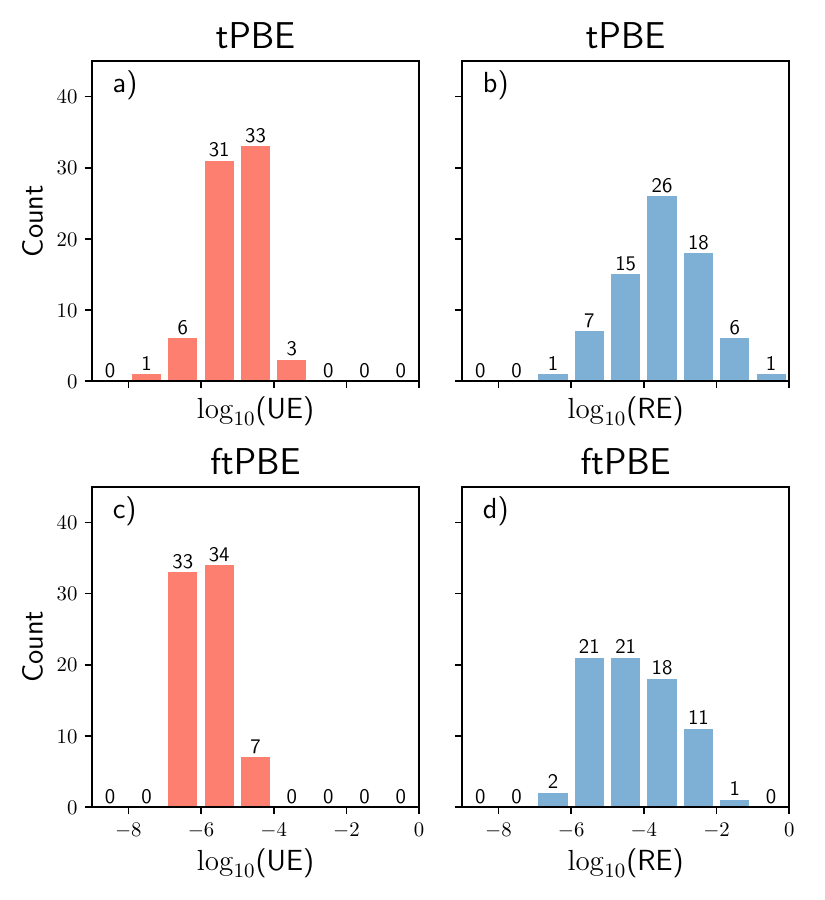}
	\caption{\label{fig:lih-hist} Distribution of the common log of the unsigned (a,c) and relative (b,d) error of analytic gradients relative to numerical gradients for all states of \ce{LiH} at various internuclear distances. The top row (a,b) is for the tPBE functional, and the bottom row (c,d) is for the ftPBE functional.}
\end{figure}

The $\log_{10}$(\glsxtrshort{ue}) and $\log_{10}$(\glsxtrshort{re}) distributions using both the tPBE and ftPBE functionals for the lowest two \textsuperscript{1}$\Sigma^+$ states of \ce{LiH} are summarized in \cref{fig:lih-hist}. We see that the majority of \glspl{ue} are below \qty{1e-4}{\hartree\per\bohr} for tPBE and below \qty{1e-5}{\hartree\per\bohr} for ftPBE. There are slightly larger \glspl{re} for both functionals in the dissociation region of the potential energy curve (Figs. S3 and S4) which are likely due to the flatness of the potential. For example, the largest \glspl{re} for tPBE and ftPBE occur on the upper \textsuperscript{1}$\Sigma^+$ state at \qty{2.7}{\angstrom} and \qty{2.8}{\angstrom} with analytic gradients of \qty{1.7532e-3} {\hartree\per\bohr} and \qty{9.0825e-5}{\hartree\per\bohr} respectively.

\Cref{tab:analytic-gradient-error} summarizes the statistical agreement between the analytic and numerical gradients for both systems and all functionals. Overall, tPBE has \gls{mue} of \qty{1.4e-5}{\hartree\per\bohr} and \qty{2.5e-5}{\hartree\per\bohr} for \ce{HeH+} and \ce{LiH} respectively; these are consistent with our previous implementations of MC-PDFT gradients. There, we saw that the analytic and numerical gradients for \ce{LiH} computed with tPBE and CMS-tPBE had \glspl{mue} of \qty{4e-5}{\hartree\per\bohr} \cite{ScottAnalytic2020} and \qty{4.9e-5}{\hartree\per\bohr} \cite{BaoAnalytic2022} respectively. 

\begin{table}
    \footnotesize
	\caption{\label{tab:analytic-gradient-error} Mean signed error (MSE), \glsxtrfull{mue}, and 
    root-mean-squared error (RMSE) of the analytic gradients relative to the numerical gradients. All values are in \unit{\hartree\per\bohr}.}
	\begin{tabular*}{\columnwidth}{@{\extracolsep{\fill}} l S[table-format=-1.1e-1] S[table-format=-1.1e-1] S[table-format=-1.1e-1] S[table-format=-1.1e-1]}
		\toprule\toprule
		& \multicolumn{2}{c}{\ce{HeH+}} & \multicolumn{2}{c}{\ce{LiH}}                              \\\cmidrule{2-3}\cmidrule{4-5}
		Functional & \mbox{tPBE}                   & \mbox{ftSVWN3}               & \mbox{tPBE} & \mbox{ftPBE} \\
		\midrule
		MSE                                & 2.2e-6                        & -1.2e-7                      & -3.8e-6     & 5.3e-7       \\
		\gls{mue}                                & 1.4e-5                        & 9.0e-7                       & 2.5e-5      & 4.4e-6       \\
		RMSE                               & 5.1e-5                        & 1.9e-6                       & 5.1e-5      & 9.1e-6       \\
		\bottomrule\bottomrule
	\end{tabular*}
\end{table}

The fully-translated functionals  have better agreement with the numerical gradients for both \ce{HeH+} and \ce{LiH} (\cref{tab:analytic-gradient-error}). This can likely be attributed to the fact that translated functionals have a discontinuity in the first derivative of the translation scheme (where values of $R$ in \cref{eq:ratio} equal 1); and since L-PDFT gradients require the second derivative, this discontinuity leads to larger numerical instabilities. The fully-translated functionals, on the other hand, use a polynomial interpolation to avoid the discontinuity (\cref{appendex:translated-functionals}), and this likely leads to more accurate gradients. 

Overall, we find good agreement between the analytic and numerical gradients at all geometries considered for both \ce{HeH+} and \ce{LiH} using both translated and fully-translated functionals.

\subsection{Formaldehyde}

Here we consider calculations with a full-valence active space of the optimized ground state ($S_0$) and the first excited state of $n\to\pi^*$ character ($S_1$) of formaldehyde. In the ground state, formaldehyde is a planar molecule with $C_{2v}$ symmetry. For the first excited state of formaldehyde, it is known that the \ce{C=O} double bond is elongated, and the molecule is no longer planar. We chose to not enforce symmetry for two reasons:  (i) One planned application of L-PDFT is to photochemistry. In photochemical simulations, one needs globally continuous potential energy surfaces, but if one enforces symmetry at high-symmetry points, one will usually get discontinuities between high-symmetry points and low-symmetry points.  So we do not enforce symmetry in photochemical simulations,  and the present paper is testing L-PDFT in that context.  (ii) The only way to include both states within the state-averaged manifold in \OpenMolcas \cite{LiManniOpenMolcas2023} is to not enforce symmetry. Not enforcing symmetry allows us to be  consistent with our prior study of formaldehyde using SA-CASSCF, SA-PDFT, and compressed multi-state PDFT (CMS-PDFT) which used the \OpenMolcas quantum chemistry package.\cite{ScottAnalytic2020,BaoAnalytic2022}. 

We define $\eta$ to be the angle between the \ce{H-C-H} plane and the \ce{C=O} bond, which is a measure of the nonplanarity of the structure. We will take as our reference the experimental ground- and excited-state geometries reported by \citet{Duncanground1974} and \citet{Jensengeometry1982} respectively.

\begin{table*}
	\caption{\label{tab:formaldehyde-geo} L-PDFT ground and first excited state bond lengths, bond angles, and out-of-plane angle ($\eta$) for formaldehyde compared with various methods. All bond lengths are in \unit{\angstrom} and all angles are in degrees. Experimental uncertainty shown in parentheses.}
	\begin{tabular*}{\textwidth}{@{\extracolsep{\fill}} l l l S[table-format=1.3(1)] S[table-format=1.3(1)] S[table-format=3.1(1)] S[table-format=2.1]}
		\toprule\toprule
		State                  & Method  & Basis Set                               & {$r_{\ce{CO}}$} & {$r_{\ce{CH}}$} & {$\theta_{\ce{HCH}}$} & {$\eta$} \\
		\midrule
		\multirow{8}{*}{$S_0$} & L-PDFT(12,10) & jun-cc-pVTZ                         & 1.210           & 1.115           & 116.4                 & 0        \\
		& SA-CASSCF(12,12) \cite{ScottAnalytic2020} & aug-cc-pVTZ & 1.214 & 1.103 & 117.3 & 0 \\
		& MC-PDFT(12,12) \cite{ScottAnalytic2020} & aug-cc-pVTZ & 1.210           & 1.114           & 116.1                 & 0        \\
		& MC-PDFT(6,5) \cite{BaoAnalytic2022}     & jun-cc-pVTZ & 1.202           & 1.112           & 115.8                 & 0        \\
		& CMS-PDFT(6,5) \cite{BaoAnalytic2022}    & jun-cc-pVTZ & 1.203           & 1.112           & 115.8                 & 0        \\
		& CASPT2(12,10) \cite{BudzakAccurate2017} & aug-cc-pVTZ & 1.209           & 1.102           & 116.1                 & 0        \\
		& \glsxtrshort{adc2} \cite{BudzakAccurate2017}        & aug-cc-pVTZ & 1.209           & 1.096           & 116.5                 & 0        \\
		& CCSD \cite{BudzakAccurate2017} & aug-cc-pVTZ & 1.201 & 1.097 & 116.4 & 0 \\
		& CCSDR(3) \cite{BudzakAccurate2017}      & aug-cc-pVTZ & 1.207           & 1.100           & 116.4                 & 0        \\
		& \glsxtrshort{cc2} \cite{BudzakAccurate2017} & aug-cc-pVTZ & 1.217 & 1.098 & 116.4 & 0 \\
		& \glsxtrshort{cc3} \cite{BudzakAccurate2017}           & aug-cc-pVTZ & 1.208           & 1.100           & 116.2                 & 0        \\
		& expt. \cite{Duncanground1974}  & &  1.207(1)          &   1.117(1)         &  116.2(1)              & 0        \\
		&                                                        \\
		\multirow{8}{*}{$S_1$} & L-PDFT(12,10)  & jun-cc-pVTZ                          & 1.328           & 1.100           & 118.1                 & 34.5       \\
		& SA-CASSCF(12,12) \cite{ScottAnalytic2020} & aug-cc-pVTZ & 1.356 & 1.079 & 118.1 & 32 \\
		& MC-PDFT(12,12) \cite{ScottAnalytic2020} & aug-cc-pVTZ & 1.323           & 1.102           & 117.6                 & 28       \\
		& MC-PDFT(6,5) \cite{BaoAnalytic2022}     & jun-cc-pVTZ & 1.333           & 1.095           & 119.9                 & 30       \\
		& CMS-PDFT(6,5) \cite{BaoAnalytic2022}    & jun-cc-pVTZ & 1.333           & 1.095           & 119.9                 & 30       \\
		& CASPT2(12,10) \cite{BudzakAccurate2017} & aug-cc-pVTZ & 1.326           & 1.090           & 118.1                 & 38       \\
		& \glsxtrshort{adc2} \cite{BudzakAccurate2017}        & aug-cc-pVTZ & 1.380           & 1.081           & 123.8                 & 19       \\
		& CCSD \cite{BudzakAccurate2017} & aug-cc-pVTZ & 1.300 & 1.087 & 118.9 & 30.9 \\
		& CCSDR(3) \cite{BudzakAccurate2017}      & aug-cc-pVTZ & 1.320           & 1.089           & 118.2                 & 37       \\
		& \glsxtrshort{cc2} \cite{BudzakAccurate2017} & aug-cc-pVTZ & 1.353 & 1.085 & 121.3 & 29.5 \\
		& \glsxtrshort{cc3} \cite{BudzakAccurate2017}           & aug-cc-pVTZ & 1.326           & 1.089           & 118.3                 & 32.7       \\
		& expt. \cite{Jensengeometry1982}         & & 1.323(3)          & 1.103(1)           & 118.1(1)                 & 34 \\
		\bottomrule\bottomrule
	\end{tabular*}
\end{table*}

The L-PDFT ground- and excited-state structural parameters are summarized in \cref{tab:formaldehyde-geo}, where they are compared to results obtained by other methods including MC-PDFT \cite{ScottAnalytic2020} and CASPT2 \cite{BudzakAccurate2017}. Both the MC-PDFT and CASPT2 calculations utilize a full-valence active space, with the MC-PDFT calculation also including two additional oxygen lone-pair orbitals. Additionally, there are results \cite{BaoAnalytic2022} from MC-PDFT and CMS-PDFT \cite{BaoCompressed2020} using a smaller (6,5) active space. This smaller active space was chosen using the ABC2 automatic active-space selection scheme \cite{BaoAutomatic2019} by setting the parameters $A$, $B$, and $C$ to 3, 2, and 0 respectively. We also include results from high-level single-reference methods computed by \citet{BudzakAccurate2017} including \gls{adc2} \cite{Dreuwalgebraic2014}, \gls{cc2} \cite{Christiansensecond1995}, \gls{cc3} \cite{Christiansensecond1995, KochCC31997}, and coupled cluster response method with single and double excitations and noniterative connected triple excitations from \gls{cc3} (CCSDR(3)) \cite{ChristiansenPerturbative1996}. 

Relative to the experimental geometry, the L-PDFT ground-state \ce{C=O} bond length differs by \qty{0.003}{\angstrom}, the \ce{C-H} bond length differs by \qty{0.002}{\angstrom}, and the \ce{H-C-H} bond angle differs by \ang{0.2} (\cref{tab:formaldehyde-geo}). For the excited state, the L-PDFT structure has a deviation of \qty{0.005}{\angstrom} for the \ce{C=O} bond length, \qty{0.003}{\angstrom} for the \ce{C-H} bond length, and \ang{0.5} for the out-of-plane dihedral ($\eta$) relative to the experimental geometry (\cref{tab:formaldehyde-geo}). The L-PDFT \ce{H-C-H} bond angle agrees with the experimental value to within \ang{0.1}. L-PDFT has the most accurate $\eta$ and \ce{H-C-H} bond angle values of any of the methods presented. 

Overall, for predicting the ground- and excited-state structures of formaldehyde, L-PDFT performs similarly to MC-PDFT with the slightly larger (12,12) active space and also similarly to the much more expensive CASPT2 method with the same active space.

\begin{table}
	\caption{\label{tab:formaldehyde-excitation} L-PDFT adiabatic and vertical excitations in \unit{\electronvolt} (not including vibration \gls{zpe}) for the first excited state of formaldehyde compared to reported values in the literature.}
	\begin{tabular*}{\columnwidth}{@{\extracolsep{\fill}}  l l S[table-format=1.2] S[table-format=1.2]}
		\toprule\toprule
		Method  & Basis Set                                             & {Adiabatic} & {Vertical} \\
		\midrule
		L-PDFT(12,10) & jun-cc-pVTZ                                       & 3.61        & 3.98       \\
		SA-CASSCF(12,12) \cite{ScottAnalytic2020} & aug-cc-pVTZ & 3.56 & 4.04		\\
		MC-PDFT(12,12) \cite{ScottAnalytic2020} & aug-cc-pVTZ             & 3.58        & 3.92       \\
		MC-PDFT(6,5) \cite{BaoAnalytic2022} & jun-cc-pVTZ                 & 3.65        & 4.07       \\
		CMS-PDFT(6,5) \cite{BaoAnalytic2022} & jun-cc-pVTZ                & 3.65        & 4.07       \\
		CASPT2(12,10) \cite{BudzakAccurate2017} & aug-cc-pVTZ             & 3.53        & 3.92       \\
		\glsxtrshort{adc2} \cite{BudzakAccurate2017} & aug-cc-pVTZ                    &             & 3.92       \\
		CCSDR(3) \cite{BudzakAccurate2017} & aug-cc-pVTZ                  &             & 3.97       \\
		\glsxtrshort{cc3} \cite{BudzakAccurate2017, Loosmountaineering2018} & aug-cc-pVTZ & 3.58         & 3.96      \\
		expt. \cite{WalzlElectron1987} &                      &             & 3.79 \\
		\bottomrule\bottomrule
	\end{tabular*}
\end{table}

\Cref{tab:formaldehyde-excitation} summarizes the adiabatic and vertical excitation energies calculated by L-PDFT and compares them with results from the literature. All values exclude the vibrational \gls{zpe}. We take the experimental vertical excitation energy measured by electron-impact spectroscopy as our reference \cite{WalzlElectron1987}. It is not possible to compare the adiabatic excitation energy to experiments without the vibrational \gls{zpe}. Instead, since the \gls{cc3} method is known to get within \qty{0.03}{\electronvolt} of the extrapolated full configuration interaction for a variety of molecules \cite{Loosmountaineering2018}, we take \gls{cc3} to be our reference for the adiabatic excitation energy. 

Like MC-PDFT, L-PDFT overestimates the vertical excitation energy with a difference of \qty{0.19}{\electronvolt} relative to the experimental value \cite{WalzlElectron1987}. However, this should be considered in the context that all methods presented in \cref{tab:formaldehyde-excitation} deviate by more than \qty{0.1}{\electronvolt} from the experimental vertical excitation energy, which has two kinds of uncertainty.  First is the uncertainty due to experimental precision, and second is uncertainty because it is only approximately true that the spectral peak in electron-impact spectroscopy corresponds to a vertical excitation energy. The L-PDFT predicted vertical excitation energy only differs from the \gls{cc3} result by \qty{0.02}{\electronvolt} and the much more expensive CASPT2 by \qty{0.06}{\electronvolt}. Additionally, L-PDFT almost exactly reproduces the \gls{cc3} adiabatic excitation energy, with a slightly higher excitation energy as compared to CASPT2. Overall, L-PDFT performs similarly to \gls{cc3}, MC-PDFT, and CASPT2 in predicting both the adiabatic and vertical excitation energies of formaldehyde.

\begin{table*}
	\caption{\label{tab:butadiene-geometry} Selected L-PDFT optimized internal coordinates for the ground and excited \textsuperscript{1}A\textsubscript{g} states of \textit{s-trans}-butadiene as compared to results from other methods. All bond lengths are in \unit{\angstrom} and all bond angles are in degrees. Experimental uncertainty shown in parentheses.}
	\begin{tabular*}{\textwidth}{@{\extracolsep{\fill}} l l l S[table-format=1.3(1)] S[table-format=1.3(1)] S[table-format=3.1(1)]}
		\toprule\toprule
		State & Method & Basis Set &  $r_{\ce{C=C}}$                                              & $r_{\ce{CC}}$                                               & $\theta_{\ce{CCC}}$ \\
		\midrule
		\multirow{6}{*}{1 \textsuperscript{1}A\textsubscript{g}} & L-PDFT(4,4) & jul-cc-pVTZ                                                       & 1.335                                                       & 1.460                                                       & 124.1                           \\
		& SA-CASSCF(4,4) \cite{ScottAnalytic2020} & aug-cc-pVTZ                           & 1.345                                                       & 1.456                                                       & 124.3                              \\
		& MC-PDFT(4,4) \cite{ScottAnalytic2020}  & aug-cc-pVTZ                            & 1.336                                                       & 1.470                                                       & 124.1                              \\
		& CASPT2(4,4) \cite{ScottAnalytic2020}   & aug-cc-pVTZ                            & 1.342                                                       & 1.454                                                       & 123.6                              \\
        & CC3 \cite{LoosReference2019} & aug-cc-pVTZ & 1.340 & 1.453 & 123.9\\
		& expt. \cite{Haugenmolecular1966}  & & 1.343(1) & 1.467(1) & 122.8(5) \\
		& \\
		\multirow{4}{*}{2 \textsuperscript{1}A\textsubscript{g}} & L-PDFT(4,4) & jul-cc-pVTZ & 1.496          & 1.399         & 124.1\\
		& SA-CASSCF(4,4) \cite{ScottAnalytic2020} & aug-cc-pVTZ & 1.489          & 1.413         & 123.2\\
		& MC-PDFT(4,4) \cite{ScottAnalytic2020} & aug-cc-pVTZ & 1.496          & 1.397         & 124.1\\
		& CASPT2(4,4) \cite{ScottAnalytic2020} & aug-cc-pVTZ & 1.488          & 1.394         & 122.1 \\
		\bottomrule\bottomrule
	\end{tabular*}
\end{table*}

\subsection{\textit{s-trans}-butadiene}

Unlike formaldehyde, \textit{s-trans}-butadiene has been shown to have a strong multireference character even in its \textsuperscript{1}A\textsubscript{g} ground state \cite{ShuDoubly2017}. \Cref{tab:butadiene-geometry} summarizes the selected optimized structural parameters of the 1 and 2 \textsuperscript{1}A\textsubscript{g} states of \textit{s-trans}-butadiene as calculated by L-PDFT and other multireference methods \cite{ScottAnalytic2020}. All active spaces are comprised of four electrons in two $\pi$ orbitals and two $\pi^*$ orbitals. Our reference for the ground-state structure is the experimental geometry from \citet{Haugenmolecular1966}, whereas for the excited-state structure we take our prior results calculated at the CASPT2(4,4) level of theory as our reference \cite{ScottAnalytic2020}. 

In general, L-PDFT performs similarly to both MC-PDFT and CASPT2 at predicting the equilibrium structures. For the ground state, L-PDFT deviates slightly from MC-PDFT for the \ce{C-C} bond length (difference of \qty{0.007}{\angstrom}), but still does better than CASPT2 and CC3. Both L-PDFT and MC-PDFT deviate the most from the experimental \ce{C=C} bond length. For the excited state, L-PDFT and MC-PDFT agree on nearly every parameter except for the \ce{C-C} bond length where they differ by only \qty{0.002}{\angstrom}. Overall, L-PDFT performs similarly to MC-PDFT, CASPT2, and CC3 at predicting the ground- and excited-state structures of the challenging \textit{s-trans}-butadiene molecule.

\begin{table}
	\caption{\label{tab:butadiene-energy} L-PDFT adiabatic and vertical excitations in \unit{\electronvolt} (not including vibrational \gls{zpe}) for the 2 \textsuperscript{1}A\textsubscript{g} state of \textit{s-trans}-butadiene compared to reported values in the literature.}
	\begin{tabular*}{\columnwidth}{@{\extracolsep{\fill}} l l S[table-format=1.2] S[table-format=1.2(1)]}
		\toprule\toprule
		Method & Basis Set & {Adiabatic} & {Vertical}                                                          \\
		\midrule
		L-PDFT(4,4)    &jul-cc-pVTZ                                                                   & 5.78        & 6.92                                                                      \\
		SA-CASSCF(4,4) \cite{ScottAnalytic2020}  & aug-cc-pVTZ                                         & 5.42        & 6.57                                                                      \\
		MC-PDFT(4,4) \cite{ScottAnalytic2020}  & aug-cc-pVTZ                                           & 5.77        & 6.91                                                                      \\
		CASPT2(4,4) \cite{ScottAnalytic2020}  & aug-cc-pVTZ                                            & 5.68        & 6.68  \\
		CCSD\footnotemark[1] \cite{ShuDoubly2017} & 6-31G** & & 7.69 \\
		MS-CASPT2(4,4)\footnotemark[1] \cite{ShuDoubly2017} & 6-31G**                                            &             & 6.69 \\
        CC3 \cite{LoosReference2019} & aug-cc-pVTZ & & 6.67\\
		\citet{WatsonExcited2012} \glsxtrshort{tbe}\footnotemark[2]                                     & &             & 6.39 \\
        \citet{LoosReference2019} \glsxtrshort{tbe}\footnotemark[2] & & & 6.50 \\
		\bottomrule\bottomrule
	\end{tabular*}
	\footnotetext[1]{Excitation energy calculated at experimental equilibrium geometry.}
	\footnotetext[2]{Theoretical best estimate.}
\end{table}

Two very-high-quality estimates of the vertical excitation energy of \textit{s-trans}-butadiene to the 2 \textsuperscript{1}A\textsubscript{g} state have been reported in the literature: \citet{WatsonExcited2012} estimated the vertical excitation energy to be 6.39 eV based on extrapolated equation-of-motion coupled-cluster calculations, and \citet{LoosReference2019} estimated it to be 6.50 eV using an extrapolated complete-basis-set FCI calculation. Multi-state CASPT2 \cite{FinleyMulti1998} using a (4,4) active space at the equilibrium geometry predicts a vertical excitation of \qty{6.69}{\electronvolt} \cite{ShuDoubly2017}, \qty{0.3}{\electronvolt} above the \gls{tbe}. \Cref{tab:butadiene-energy} summarizes the adiabatic and vertical excitation energies of L-PDFT as compared to the previously computed SA-CASSCF, MC-PDFT, and CASPT2 energies using a two-state model space with a (4,4) active space \cite{ScottAnalytic2020} as well as the CC3 vertical . Although the L-PDFT adiabatic excitation energy differs by only \qty{0.1}{\electronvolt} from the CASPT2 predicted adiabatic excitation energy and by only \qty{0.01}{\electronvolt} from MC-PDFT, it overestimates the vertical excitation energy by more than \qty{0.4}{\electronvolt} as compared to the best available estimates.

\subsection{Phenol}

The photochemistry of phenol has been extensively studied as it is a prototype of the $^1\pi\sigma^*$ motif which is common in a variety of biomolecules and aromatic compounds \cite{SobolewskiExcited2002, Ashfoldrole2006, Devineultraviolet2008, AshfoldExploring2008, LimPhotodissociation2009, XuDiabatic2013, Zhuimproved2016, ZhangElectronic2018, ZhangFull2019}. In the original L-PDFT paper, we studied the \ce{O-H} photodissociation potential energy surface and found that L-PDFT was able to correctly model the potential energy surface near the conical intersection, whereas MC-PDFT surfaces unphysically crossed \cite{HennefarthLinearized2023}. Our active space in this paper is the same as in our prior studies of phenol \cite{HennefarthLinearized2023, BaoAnalytic2022, BaoCompressed2020} consisting of $3(\pi,\pi^*)$, the p\textsubscript{z} of \ce{O}, and the \ce{C-O} and \ce{O-H} $\sigma$ and $\sigma^*$ orbitals.

\begin{table*}
	\caption{\label{tab:phenol-geometry} L-PDFT selected internal coordinates for the ground- and first excited-state of phenol compared with various methods. All bond lengths are in \unit{\angstrom} and all angles are in degrees. The experimental uncertainty is shown in parentheses.}
	\begin{tabular*}{\textwidth}{@{\extracolsep{\fill}} l l l S[table-format=1.4] S[table-format=1.4(1)] S[table-format=1.4(1)] S[table-format=3.1(1)] S[table-format=2.1]}
		\toprule\toprule
		State                  & Method & Basis Set & {Avg. $r_{\ce{CC}}$} & {$r_{\ce{CO}}$} & {$r_{\ce{OH}}$} & {$\theta_{\ce{COH}}$} & {$\tau_{\ce{CCOH}}$} \\
		\midrule
		\multirow{6}{*}{$S_0$} & L-PDFT(12,11) & jul-cc-pVDZ                                                                        & 1.400                & 1.369           & 0.962           & 109.4                 & 0.0                  \\
		& SA-CASSCF(12,11) \cite{BaoAnalytic2022} & jul-cc-pVDZ                                              & 1.399                & 1.384           & 0.966           & 109.3                 & 0.0                  \\
		& MC-PDFT(12,11) \cite{BaoAnalytic2022} & jul-cc-pVDZ                                                & 1.401                & 1.370           & 0.964           & 109.2                 & 0.0                  \\
		& CMS-PDFT(12,11) \cite{BaoAnalytic2022} & jul-cc-pVDZ                                               & 1.398                & 1.367           & 0.966           & 109.2                 & 0.0                  \\
		& semiemp. fit \cite{Zhuimproved2016}                                           & & 1.395                & 1.382           & 0.965           & 108.5                 & 0.0                  \\
		& expt. \cite{LarsenMicrowave1979} &                                              & 1.393                & 1.375(5)  & 0.957(6)         & 108.8(4)                 & 0.0                  \\
		&                                                                                                                                                                                          \\
		\multirow{6}{*}{$S_1$} & L-PDFT(12,11) & jul-cc-pVDZ                                                                        & 1.431                & 1.346           & 0.964           & 109.4                 & 0.0                  \\
		& SA-CASSCF(12,11) \cite{BaoAnalytic2022} & jul-cc-pVDZ                                              & 1.434                & 1.379           & 0.960           & 109.3                 & 0.0                  \\
		& MC-PDFT(12,11) \cite{BaoAnalytic2022} & jul-cc-pVDZ                                                & 1.429                & 1.337           & 0.975           & 108.2                 & 14.3                 \\
		& CMS-PDFT(12,11) \cite{BaoAnalytic2022} & jul-cc-pVDZ                                               & 1.435                & 1.362           & 0.962           & 109.2                 & 0.0                  \\
		& semiemp. fit \cite{Zhuimproved2016}                                            & & 1.427                & 1.367           & 0.963           & 108.7                 & 0.0                  \\
		& expt. \cite{SpangenbergS12003} &                                                & 1.423                & 1.356           & 0.992           & 108.8                 & 0.0 \\
		\bottomrule\bottomrule
	\end{tabular*}
\end{table*}

Selected L-PDFT optimized ground- and first excited-state internal coordinates of phenol are presented in \cref{tab:phenol-geometry} and are compared with results from other, similar methods. Our reference for the ground- and excited-state geometries are the experimental structures determined by \citet{LarsenMicrowave1979} and \citet{SpangenbergS12003} respectively. All of the PDFT and CASSCF methods in \cref{tab:phenol-geometry} use the same (12,11) active space. We also include results from a high-level semiempirical fit that was designed to replicate the \gls{zpe}-inclusive experimental adiabatic excitation energy \cite{Zhuimproved2016}. All of the methods in the table predict relatively similar ground-state geometries, in line with the experimentally determined geometry \cite{LarsenMicrowave1979}. Of interest is that the excited-state geometry optimized with MC-PDFT is nonplanar, with a substantial \ce{C-C-O-H} dihedral of \ang{14.3} \cite{BaoAnalytic2022}. It was noted that CMS-PDFT does not suffer from this incorrect nonplanarity because it correctly incorporates the state interaction between $S_0$ and $S_1$. L-PDFT also correctly predicts a planar excited-state geometry, in agreement with CMS-PDFT \cite{BaoAnalytic2022} and experimental results \cite{SpangenbergS12003}. This confirms that L-PDFT accounts for the state interaction as well as CMS-PDFT does. Overall, L-PDFT performs better than MC-PDFT in accurately predicting the first excited-state geometry of phenol, and the results are similar to those for CMS-PDFT and other high-level methods.

\begin{table}
	\caption{\label{tab:phenol-excitation} L-PDFT adiabatic and vertical excitations in \unit{\electronvolt} (not including vibration \gls{zpe}) for the first excited state of phenol compared to reported values in the literature.}
	\begin{tabular*}{\columnwidth}{@{\extracolsep{\fill}} l l S[table-format=1.2] S[table-format=1.2]}
		\toprule\toprule
		Method & Basis Set                                                                       & {Adiabatic} & {Vertical}             \\
		\midrule
		L-PDFT(12,11) & jul-cc-pVDZ                                                                & 4.85      & 5.03                 \\
		SA-CASSCF(12,11) \cite{BaoAnalytic2022} & jul-cc-pVDZ                                      & 4.73      & 4.93                 \\
		MC-PDFT(12,11) \cite{BaoAnalytic2022} & jul-cc-pVDZ                                         & 4.83      & 5.03                 \\
		CMS-PDFT(12,11) \cite{BaoAnalytic2022} & jul-cc-pVDZ                                       & 4.72      & 4.93                 \\
		CASPT2(8,8)\footnotemark[1] \cite{Granuccitheoretical2000} & cc-pVDZ & 4.36      & 4.64                 \\
		CASPT2(10,10)\footnotemark[1] \cite{DixonTunnelling2011} & aug(O)-AVTZ\footnotemark[2]  & 4.37      & 4.52                 \\
		\glsxtrshort{cc2}\footnotemark[3] \cite{PinoExcited2010} & aug-cc-pVDZ                & 4.67      & 4.86                 \\
		\glsxtrshort{mrci}(10,9)\footnotemark[1]\footnotemark[4] \cite{VieuxmaireAb2008} & aug-cc-pVDZ           & 4.82      & 4.75 \\
		semiemp. fit \cite{Zhuimproved2016} &                                          & 4.66      & 4.83 \\
		\bottomrule\bottomrule
	\end{tabular*}
	\footnotetext[1]{Excitations calculated at the CASSCF optimized geometry.}
	\footnotetext[2]{Modified aug-cc-pVTZ basis set with extra even tempered sets of $s$ and $p$ diffuse functions on the oxygen atom.}
	\footnotetext[3]{Ground state optimized with \glsxtrshort{mp2} and excited state optimized with \glsxtrshort{cc2}.}
	\footnotetext[4]{It is possible for the vertical excitation to be lower than the adiabatic excitation when the excitations are computed at geometries optimized at a different level of theory.}
\end{table}

\begin{table*}
    \footnotesize
	\caption{\label{tab:cytosine-ground-bond} Selected L-PDFT ground state cytosine bond lengths (in \unit{\angstrom}) compared with similar methods and experimental quantities. Atoms are labeled according to \cref{fig:cytosine-opt}.}
	\begin{tabular*}{\textwidth}{@{\extracolsep{\fill}} l l S[table-format=1.3] S[table-format=1.3] S[table-format=1.3] S[table-format=1.3] S[table-format=1.3] S[table-format=1.3] S[table-format=1.3] S[table-format=1.3] c}
		\toprule\toprule
		Method & Basis Set                                   & {C1-N2} & {N2-C3} & {C3-C4} & {C4-C5} & {C5-N7} & {C5-N6} & {N6-C1} & {C1-O8} & \glsxtrshort{mud}\footnote{Mean unsigned deviation from experiment.} \\
		\midrule
		L-PDFT(14,10) & jul-cc-pVTZ                             & 1.438   & 1.344   & 1.361   & 1.434   & 1.352   & 1.327   & 1.363   & 1.218   & 0.02                                                   \\
		SA-CASSCF(14,10) \cite{ScottAnalytic2020} & aug-cc-pVTZ & 1.391   & 1.354   & 1.346   & 1.446   & 1.350   & 1.291   & 1.391   & 1.296   & 0.02                                                   \\
		MC-PDFT(14,10) \cite{ScottAnalytic2020} & aug-cc-pVTZ  & 1.440   & 1.340   & 1.358   & 1.436   & 1.351   & 1.326   & 1.364   & 1.215   & 0.02                                                   \\
		CCSD \cite{FogarasiRelative2002} & TZP      & 1.416   & 1.360   & 1.353   & 1.446   & 1.357   & 1.313   & 1.379   & 1.214   & 0.02                                                   \\
		CASPT2(14,10) \cite{ScottAnalytic2020} & 6-311G+(2df)   & 1.420   & 1.362   & 1.357   & 1.443   & 1.360   & 1.320   & 1.378   & 1.222   & 0.02                                                   \\
		expt. \cite{BarkerCrystal1964} &           & 1.374   & 1.357   & 1.342   & 1.424   & 1.330   & 1.337   & 1.364   & 1.234   & \\
		\bottomrule\bottomrule
	\end{tabular*}
\end{table*}

\begin{table*}
	\caption{\label{tab:cytosine-ground-angle} Selected L-PDFT ground state cytosine bond angles (in degrees) compared with similar methods and experimental quantities. Atoms are labeled according to \cref{fig:cytosine-opt}.}
	\begin{tabular*}{\textwidth}{@{\extracolsep{\fill}} l S[table-format=3.1] S[table-format=3.1] S[table-format=3.1] S[table-format=3.1] S[table-format=3.1] S[table-format=3.1] S[table-format=3.1] S[table-format=3.1] S[table-format=3.1] S[table-format=3.1] c}
		\toprule\toprule
		Method\footnote{Methods are the same as in \cref{tab:cytosine-ground-bond}.}                                    & {$\theta$\textsubscript{6-1-2}} & {$\theta$\textsubscript{5-6-1}} & {$\theta$\textsubscript{4-5-6}} & {$\theta$\textsubscript{3-2-1}} & {$\theta$\textsubscript{3-4-5}} & {$\theta$\textsubscript{4-3-2}} & {$\theta$\textsubscript{8-1-2}} & {$\theta$\textsubscript{8-1-6}} & {$\theta$\textsubscript{7-5-6}} & {$\theta$\textsubscript{7-5-4}} & \glsxtrshort{mud}\footnote{Mean unsigned deviation from experiment.} \\
		\midrule
		L-PDFT(14,10)                             & 116.0                           & 120.4                           & 124.2                           & 123.2                           & 115.8                           & 120.4                           & 117.3                           & 126.6                           & 116.2                           & 119.6                           & 1.6                                                    \\
		SA-CASSCF(14,10) \cite{ScottAnalytic2020} & 119.3                           & 119.3                           & 120.6                           & 122.3                           & 118.4                           & 120.3                           & 120.0                           & 120.8                           & 118.0                           & 121.7                           & 0.9                                                    \\
		MC-PDFT(14,10) \cite{ScottAnalytic2020}   & 115.8                           & 120.5                           & 124.0                           & 123.4                           & 115.7                           & 119.6                           & 117.5                           & 117.7                           & 116.4                           & 119.5                           & 1.6                                                    \\
		CCSD \cite{FogarasiRelative2002}          & 119.7                           & 120.1                           & 119.5                           & 121.9                           & 118.8                           & 120.2                           & 120.9                           & 119.6                           & 119.2                           & 121.5                           & 1.3                                                    \\
		CASPT2(14,10) \cite{ScottAnalytic2020}    & 119.9                           & 120.1                           & 119.3                           & 121.7                           & 118.9                           & 120.3                           & 120.9                           & 119.4                           & 119.5                           & 121.4                           & 1.4                                                    \\
		expt. \cite{BarkerCrystal1964}            & 119.1                           & 119.9                           & 122.0                           & 122.7                           & 117.3                           & 120.1                           & 119.8                           & 122.2                           & 118.2                           & 119.9                           & \\
		\bottomrule\bottomrule
	\end{tabular*}
\end{table*}

\begin{table*}
    \footnotesize
	\caption{\label{tab:cytosine-excited-bond} Selected L-PDFT 2 \textsuperscript{1}A\textsuperscript{$\prime$} excited state cytosine bond lengths (in \unit{\angstrom}) compared with similar methods and experimental quantities. Atoms are labeled according to \cref{fig:cytosine-opt}.}
	\begin{tabular*}{\textwidth}{@{\extracolsep{\fill}} l l S[table-format=1.3] S[table-format=1.3] S[table-format=1.3] S[table-format=1.3] S[table-format=1.3] S[table-format=1.3] S[table-format=1.3] S[table-format=1.3] c}
		\toprule\toprule
		Method & Basis Set                                   & {C1-N2} & {N2-C3} & {C3-C4} & {C4-C5} & {C5-N7} & {C5-N6} & {N6-C1} & {C1-O8} & \glsxtrshort{mud}\footnote{Mean unsigned deviation from MS-CASPT2.} \\
		\midrule
		L-PDFT(14,10) & jul-cc-pVTZ                            & 1.398   & 1.390   & 1.415   & 1.376   & 1.361   & 1.406   & 1.313   & 1.271   & 0.02                                                  \\
		SA-CASSCF(14,10) \cite{ScottAnalytic2020} & aug-cc-pVTZ & 1.358   & 1.381   & 1.424   & 1.365   & 1.372   & 1.417   & 1.274   & 1.328   & 0.04                                                  \\
		MC-PDFT(14,10) \cite{ScottAnalytic2020} & aug-cc-pVTZ  & 1.429   & 1.382   & 1.416   & 1.385   & 1.369   & 1.416   & 1.342   & 1.236   & 0.02                                                  \\
		CASPT2(14,10) \cite{ScottAnalytic2020} & 6-311G+(2df)   & 1.453   & 1.301   & 1.394   & 1.444   & 1.350   & 1.320   & 1.374   & 1.210   & 0.05                                                  \\
		MS-CASPT2(8,7) \cite{NakayamaQuantum2014} & Sapporo-DZP & 1.410   & 1.376   & 1.430   & 1.391   & 1.397   & 1.351   & 1.325   & 1.276   &                                                       \\
		\bottomrule\bottomrule
	\end{tabular*}
\end{table*}

\begin{table*}
	\caption{\label{tab:cytosine-excited-angle} Selected L-PDFT 2 \textsuperscript{1}A\textsuperscript{$\prime$} excited state cytosine bond angles (in degrees) compared with similar methods and experimental quantities. Atoms are labeled according to \cref{fig:cytosine-opt}.}
	\begin{tabular*}{\textwidth}{@{\extracolsep{\fill}} l S[table-format=3.1] S[table-format=3.1] S[table-format=3.1] S[table-format=3.1] S[table-format=3.1] S[table-format=3.1] S[table-format=3.1] S[table-format=3.1] S[table-format=3.1] S[table-format=3.1] c}
		\toprule\toprule
		Method\footnote{Methods are the same as in \cref{tab:cytosine-excited-bond}.}                                    & {$\theta$\textsubscript{6-1-2}} & {$\theta$\textsubscript{5-6-1}} & {$\theta$\textsubscript{4-5-6}} & {$\theta$\textsubscript{3-2-1}} & {$\theta$\textsubscript{3-4-5}} & {$\theta$\textsubscript{4-3-2}} & {$\theta$\textsubscript{8-1-2}} & {$\theta$\textsubscript{8-1-6}} & {$\theta$\textsubscript{7-5-6}} & {$\theta$\textsubscript{7-5-4}} & \glsxtrshort{mud}\footnote{Mean unsigned deviation from MS-CASPT2.} \\
		\midrule
		L-PDFT(14,10)                             & 122.7                           & 117.1                           & 122.6                           & 121.7                           & 119.9                           & 115.9                           & 114.1                           & 123.2                           & 111.9                           & 125.5                           & 1.1                                                   \\
		SA-CASSCF(14,10) \cite{ScottAnalytic2020} & 126.6                           & 116.1                           & 121.6                           & 119.8                           & 119.6                           & 116.4                           & 112.1                           & 121.2                           & 112.7                           & 125.8                           & 1.6                                                   \\
		MC-PDFT(14,10) \cite{ScottAnalytic2020}   & 121.4                           & 118.1                           & 121.7                           & 121.1                           & 120.4                           & 117.4                           & 115.7                           & 122.9                           & 112.2                           & 126.2                           & 1.9                                                   \\
		CASPT2(14,10) \cite{ScottAnalytic2020}    & 118.1                           & 119.1                           & 122.8                           & 123.4                           & 118.4                           & 118.3                           & 116.7                           & 125.2                           & 119.0                           & 118.3                           & 3.0                                                   \\
		MS-CASPT2(8,7) \cite{NakayamaQuantum2014} & 123.0                           & 115.8                           & 123.4                           & 122.7                           & 119.6                           & 115.4                           & 111.8                           & 125.2                           & 112.8                           & 123.7                                                                                   \\
		\bottomrule\bottomrule
	\end{tabular*}
\end{table*}

\Cref{tab:phenol-excitation} contains a summary of the vertical and adiabatic excitation energies of phenol as obtained by L-PDFT and other methods. As in our previous work \cite{BaoAnalytic2022}, we take as our reference the high-level semiempirical fit by \citet{Zhuimproved2016} for both excitation energies. We also include results from \gls{cc2} \cite{PinoExcited2010} (for which the ground-state geometries are optimized at the \gls{mp2} \cite{MoellerNote1934, HeadGordonMP21988}), internally contracted \gls{mrci} \cite{KnowlesInternally1992} based on a CASSCF(10,9) wave function \cite{VieuxmaireAb2008}, and CASPT2 using a (8,8) \cite{Granuccitheoretical2000} and (10,10) \cite{DixonTunnelling2011} active space. Both the \gls{mrci} and CASPT2 results were computed at their respective reference CASSCF optimized geometries. 

L-PDFT performs similarly to the previously reported MC-PDFT results \cite{BaoAnalytic2022} and overestimates the vertical and adiabatic excitation energy relative to the reference by about \qty{0.2}{\electronvolt}. Comparatively, CASPT2 using both an (8,8) \cite{Granuccitheoretical2000} and (10,10) \cite{DixonTunnelling2011} active space underestimates the vertical and adiabatic excitation energy by more than \qty{0.2}{\electronvolt}. The relative difference between the L-PDFT vertical and adiabatic excitation energies is \qty{0.18}{\electronvolt} which is similar to the relative difference of \qty{0.17}{\electronvolt} predicted by the semiempirical fit \cite{Zhuimproved2016}.

\subsection{Cytosine}

\begin{figure}
	\centering
	\includegraphics[width=3in]{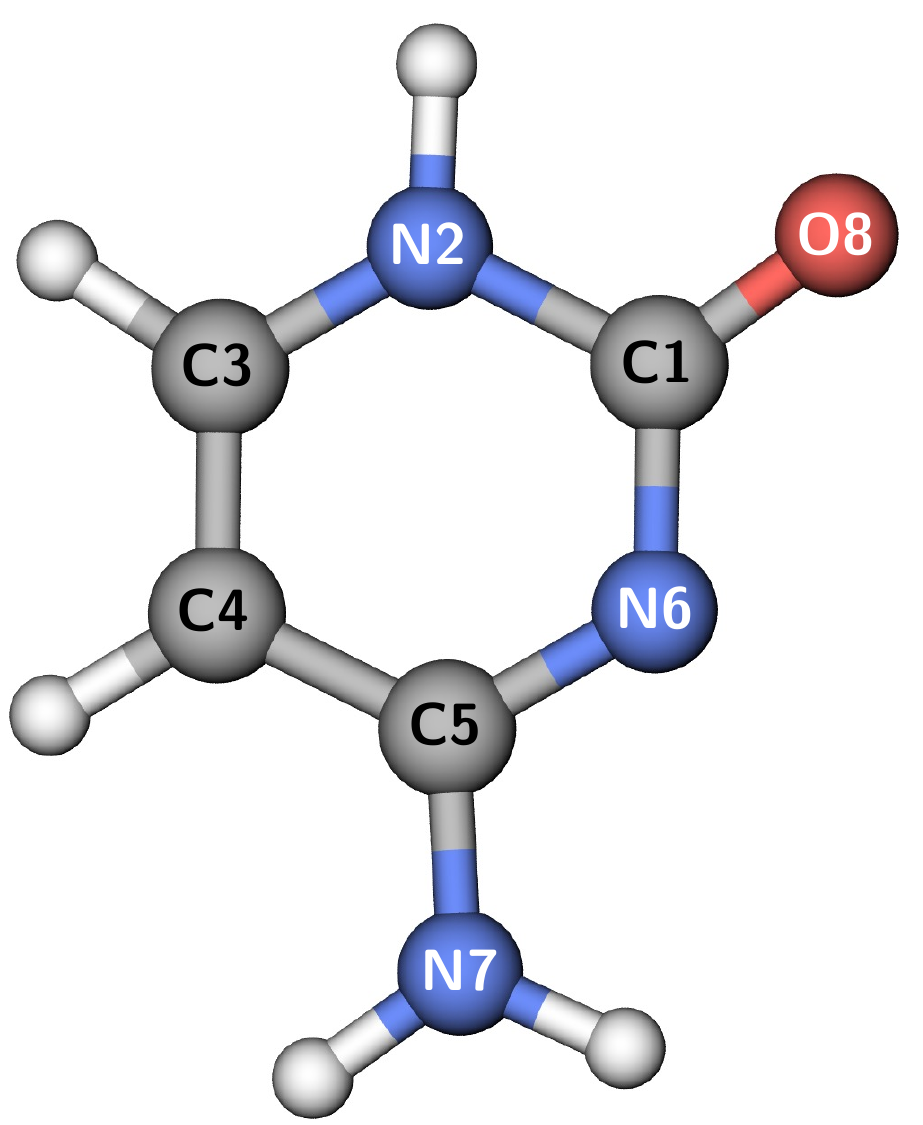}
	\caption{\label{fig:cytosine-opt} Ground-state geometry of cytosine optimized with L-PDFT.}
\end{figure}

Finally, we report the optimized ground- and excited-state geometries and the adiabatic and vertical excitation energies of the nucleobase cytosine (\cref{fig:cytosine-opt}). Previous studies have shown that the (14,10) active space composed of five $\pi$, two lone-pair, and three $\pi^*$ orbitals is sufficient for studying the low-lying excited states of cytosine \cite{MerchanUnified2006, GonzalezVazqueztime2010, NakayamaPhotophysics2013}.

For the ground state, we take the experimental structure determined by \citet{BarkerCrystal1964} as our reference. \Cref{tab:cytosine-ground-bond,tab:cytosine-ground-angle} contain selected optimized bond lengths and angles respectively obtained from L-PDFT and other methods. All methods presented in \cref{tab:cytosine-ground-bond} perform similarly with a \gls{mud} of \qty{0.02}{\angstrom} relative to the experimental bond lengths. L-PDFT performs identically to MC-PDFT for determining the cytosine bond angles with an \gls{mud} of \ang{1.6} for both methods. 

Due to the lack of experimental data for the 2 \textsuperscript{1}A\textsuperscript{$\prime$} relaxed geometry, we take the MS-CASPT2 with an (8,7) active space from the study by \citet{NakayamaQuantum2014} as our reference. \Cref{tab:cytosine-excited-bond,tab:cytosine-excited-angle} compares the same selected 2 \textsuperscript{1}A\textsuperscript{$\prime$} optimized bond lengths and bond angles for L-PDFT and other methods. All methods give similar optimized bond lengths for the excited state with L-PDFT and MC-PDFT both having a \gls{mud} of \qty{0.02}{\angstrom}. CASPT2 and SA-CASSCF have relatively larger \glspl{mud} of \qty{0.04}{\angstrom} and \qty{0.05}{\angstrom} respectively. L-PDFT and MC-PDFT predict similar bond angles of the excited state, with L-PDFT being slightly closer to the MS-CASPT2 results with a \gls{mud} of \ang{1.1}. CASPT2 has the widest difference in the bond angles, and it overestimates the 7-5-6 bond angle by about \ang{7} as compared to all of the other methods.

\begin{table}
	\caption{\label{tab:cytosine-energy} L-PDFT adiabatic and vertical excitations in \unit{\electronvolt} (not including vibration \gls{zpe}) for the 2 \textsuperscript{1}A\textsuperscript{$\prime$} state of cytosine compared to reported values in the literature.}
	\begin{tabular*}{\columnwidth}{@{\extracolsep{\fill}} l l S[table-format=1.2] S[table-format=1.2]}
		\toprule\toprule
		Method & Basis Set                                                                                                                                                                    & {Adiabatic} & {Vertical} \\
		\midrule
		L-PDFT(14,10) & jul-cc-pVTZ                                                                                                                                                             & 4.05        & 4.42       \\
		MC-PDFT(14,10) \cite{ScottAnalytic2020} & aug-cc-pVTZ                                                                                                                                    & 4.08        & 4.45       \\
		\glsxtrshort{cc2} \cite{ChernevaExperimental2023} & aug-cc-pVDZ                                                                                                                                        &             & 4.54       \\
		MS-CASPT2(12,9)\footnote{Ground-state geometry optimized with \glsxtrshort{mp2} and excited-state geometry optimized with MS-CASPT2(8,7).} \cite{NakayamaQuantum2014} & Sapporo-DZP & 3.98        & 4.48       \\
		expt. \cite{Abouaftriplet2004} &                                                                                                                                            &             & 4.65 \\
		\bottomrule\bottomrule
	\end{tabular*}
\end{table}

\Cref{tab:cytosine-energy} summarizes the adiabatic and vertical excitation energies for the 2 \textsuperscript{1}A\textsuperscript{$\prime$} state of cytosine computed at various levels of theory. We take the experimental vertical excitation from \citet{Abouaftriplet2004}, and the MS-CASPT2(12,9) result (ground-state geometry optimized with \gls{mp2} and excited-state geometry optimized with MS-CASPT2(8,7)) as the reference for the adiabatic excitation energy \citet{NakayamaQuantum2014}.  All methods, except \gls{cc2}, perform similarly with L-PDFT underestimating the reference vertical excitation energy the most. \Gls{cc2} gets the closest to the experimental vertical excitation energy, differing only by \qty{0.11}{\electronvolt}. Both L-PDFT and MC-PDFT get within \qty{0.1}{\electronvolt} of the MS-CASPT2 adiabatic excitation energy. Overall, L-PDFT differs from MC-PDFT by only \qty{0.03}{\electronvolt} for both the adiabatic and vertical excitation energies.

\section{Conclusion}

We presented the derivation and implementation of analytic nuclear gradients for L-PDFT calculations based on SA-CASSCF wave functions. Because the final L-PDFT wave function is not fully variational with respect to all its parameters, we used a Lagrangian method similar to that used previously for SA-CASSCF and MC-PDFT analytic gradients. As in SA-CASSCF, we assumed equal weights to exclude the model states from the response equations. We then implemented the gradients in \PySCFforge, which is a library of \PySCF extensions, and we showed that they agree with numerical gradients for both \ce{HeH+} and \ce{LiH} using both translated and fully-translated functionals. We showed the utility of the L-PDFT analytic gradients by optimizing the ground and first excited singlet states of formaldehyde, \textit{s-trans}-butadiene, phenol, and the nucleobase cytosine. Whereas MC-PDFT predicts a nonplanar first excited state of phenol, we showed that L-PDFT correctly predicts a planar structure. Additionally, we computed the vertical and adiabatic excitation energy for each molecule and saw that L-PDFT performs similarly for excitation energies to MC-PDFT and other high-level multireference methods like CASPT2.
 
These results are consistent with our prior study and benchmarking of L-PDFT for calculating vertical excitation energies and for modeling potential energy surfaces \cite{HennefarthLinearized2023a, HennefarthLinearized2023}. Specifically, L-PDFT correctly models the potential energy surfaces near conical intersections and locally avoided crossings whereas MC-PDFT is inaccurate \cite{HennefarthLinearized2023}. The results are especially encouraging because of the low cost of L-PDFT relative to MS-CASPT2 and \gls{mrci}. We conclude that L-PDFT is promising new tool for studying exited-state geometries, both vertical and adiabatic energies, photochemical reactions, and electronically nonadiabatic dynamics.

\section*{Supplementary Material}

See the supplementary material associated with this article for the analytic and numerical gradients of \ce{HeH+} and \ce{LiH} at each geometry and L-PDFT optimized structures with their corresponding energies.

\begin{acknowledgments}
	This work was supported in part by the National Science Foundation under Grant
	No. CHE-2054723. M.R. Hennefarth acknowledges support by the National Science
	Foundation Graduate Research Fellowship under Grant No. 2140001. We also
	acknowledge the University of Chicago’s Research Computing Center for their
	support of this work. Any opinion, findings, and conclusions or recommendations
	expressed in this material are those of the author(s) and do not necessarily
	reflect the views of the National Science Foundation.
\end{acknowledgments}

\appendix

\section{The Translated and Fully-Translated On-Top Functionals \label{appendex:translated-functionals}}

Current generation on-top functionals include translated \cite{LiManniMulticonfiguration2014} or fully-translated \cite{CarlsonMulticonfiguration2015} \gls{ks} local-spin density approximations or generalized gradient approximations functionals. These on-top functionals are defined such that
\begin{equation}
	\ot\bqty{\vrho} = \xc\bqty{\vtrho}
\end{equation}
where $\xc$ is a \gls{ks} exchange-correlation functional and $\vtrho$ are the collective translated (or fully-translated) spin-density variables and their gradients.
\begin{equation}
  \transpose{\vtrho} = \begin{bmatrix}
    \rhoa & \rhob & \sigmaaa & \sigmaab & \sigmabb 
  \end{bmatrix} 
\end{equation}
Here, $\rhoa$ and $\rhob$ are effective spin densities, with primes denoting the gradient with respect to electron coordinate; and $\sigmaaa$, $\sigmaab$, and $\sigmabb$ being the inner product of the effective spin density gradients.
\begin{subequations}
\begin{align}
  \sigmaaa &= \rhopa \cdot \rhopa \\ 
  \sigmaab &= \rhopa \cdot \rhopb \\
  \sigmabb &= \rhopb \cdot \rhopb
\end{align} 
\end{subequations}

For translated functionals, the following mapping is used to generate the effective spin densities and their gradients from the wave function's density and on-top pair density:
\begin{equation}
	\rhoa = \frac{\rho}{2}\pqty{1+\tzeta}
\end{equation}
\begin{equation}
	\rhob = \frac{\rho}{2}\pqty{1-\tzeta}
\end{equation}
\begin{equation}
	\rhopa = \frac{\rhop}{2}\pqty{1+\tzeta}
\end{equation}
\begin{equation}
	\rhopb = \frac{\rhop}{2}\pqty{1-\tzeta}
\end{equation}
where $\tzeta$ is given by
\begin{equation}
	\tzeta = \begin{cases}
		\sqrt{1-R} & R \leq 1 \\
		0          & R > 1
	\end{cases}
\end{equation}
and $R$ is proportional to the ratio of the on-top density to the density.
\begin{equation}
	\label{eq:ratio}
	R = \frac{4\Pi}{\rho^2}
\end{equation}
Both $\tzeta$ and $R$ are functions of $\vb{r}$. Functionals translated by this scheme are simply known as `translated' functionals and are given the prefix `t' \cite{LiManniMulticonfiguration2014}.

The above translation scheme has a discontinuity in its first derivative at $R = 1$. The fully-translated scheme \cite{CarlsonMulticonfiguration2015} fixes this by using a polynomial interpolation to smooth out the discontinuity in the region of $R$ close to 1 as
\begin{equation}
	\rhoa = \frac{\rho}{2}\pqty{1+\ftzeta}
\end{equation}
\begin{equation}
	\rhob = \frac{\rho}{2}\pqty{1-\ftzeta}
\end{equation}
\begin{equation}
	\rhopa = \frac{\rhop}{2}\pqty{1+\ftzeta} + \frac{\rho}{2}\ftzetap
\end{equation}
\begin{equation}
	\rhopb = \frac{\rhop}{2}\pqty{1-\ftzeta} - \frac{\rho}{2}\ftzetap
\end{equation}
where $\ftzeta$ is 
\begin{equation}
	\ftzeta = \begin{cases}
		\sqrt{1-R} & R < R_0             \\
		P(R)       & R_0 \leq R \leq R_1 \\
		0          & R > R_1
	\end{cases}
\end{equation}
and $P$ is the polynomial interpolation from $R_0$ to $R_1$.
\begin{equation}
	P = P(R) = A\pqty{R-R_1}^5 + B\pqty{R-R_1}^4 + C\pqty{R-R_1}^3
\end{equation}
The gradient of $\ftzeta$ is given by
\begin{equation}
	\ftzetap = \begin{cases}
		-\frac{R^\prime}{2\ftzeta} & R < R_0             \\
		R^\prime P^{(1)}           & R_0 \leq R \leq R_1 \\
		0                          & R > R_1
	\end{cases}
\end{equation}
where $R^\prime$ is written as
\begin{equation}
	R^\prime = \frac{4\Pip}{\rho^2} - \frac{8\Pi\rhop}{\rho^3} = R\pqty{\frac{\Pip}{\Pi} - \frac{2\rhop}{\rho}}
\end{equation}
and $P^{(1)}$ is the first derivative of $P$ with respect to $R$.
\begin{equation}
	P^{(1)} = 5A\pqty{R-R_1}^4 + 4B\pqty{R-R_1}^3 + 3C\pqty{R-R_1}^2
\end{equation}
The parameters $A, B, C, R_0,$ and $R_1$ are given in \cref{appendix-tab:ft-parameters}

\begin{table}
	\caption{\label{appendix-tab:ft-parameters} Parameter values for the fully-translated scheme.}
	\begin{tabular*}{\columnwidth}{@{\extracolsep{\fill}} l S}
		\toprule\toprule
		Parameter & {Value}       \\
		\midrule
		$R_0$     & 0.9           \\
		$R_1$     & 1.15          \\
		$A$       & -475.60656009 \\
		$B$       & -379.47331922 \\
		$C$       & -85.38149682 \\
		\bottomrule\bottomrule
	\end{tabular*}
\end{table}

$\tzeta$ can be considered a special case of $\ftzeta$ with $R_0 = R_1 = 1$. Then, $\zeta$ can be used to denote either $\tzeta$ or $\ftzeta$ with it being clear from the context which form is being used. Consequently, the main difference between translated and fully-translated functionals is the form of $\rhopa$ and $\rhopb$.

\section{SA-CASSCF Hessian in the L-PDFT Eigenstate Basis \label{appendex:modified-casscf-hessian}}

The CI portion of the SA-CASSCF Hessian matrix, $\hess^{E^\mathrm{SA}}_{\vb{P}^\perp \vb{P}^\perp}$, can be expressed either in the SA-CASSCF or L-PDFT eigenstate basis. We define $\Theta^{MN}_{IJ}$ and $\tilde{\Theta}^{MN}_{\Lambda \Gamma}$ as the elements of the CI Hessian in the SA-CASSCF and L-PDFT eigenstate bases respectively.
\begin{equation}
	\Theta^{MN}_{IJ} = \pdv{E^\mathrm{SA}}{P^I_M}{P^J_N}
\end{equation}
\begin{equation}
	\tilde{\Theta}^{MN}_{\Lambda \Gamma} = \pdv{E^\mathrm{SA}}{P^\Lambda_M}{P^\Gamma_N}
\end{equation}
From Eq. 30 of Ref. \citenum{StaalringAnalytical2001}, the elements of $\hess^{E^\mathrm{SA}}_{\vb{P}^\perp \vb{P}^\perp}$ in the SA-CASSCF eigenstate basis can be written as
\begin{equation}
	\Theta^{MN}_{IJ} = \frac{2\delta_{IJ}}{n^\mathrm{SA}}\mel{M}{\pqty{\hat{H}^\mathrm{el}-E^\mathrm{CAS}_I}}{N}
\end{equation}
where $n^\mathrm{SA}$ are the number of states in the model space, $\hat{H}^\mathrm{el}$ is the real electronic Hamiltonian, and $E^\mathrm{CAS}_I$ is the CASSCF energy for state $\ket{I}$.

In our implementation, we evaluate $\hess^{E^\mathrm{SA}}_{\vb{P}^\perp \vb{P}^\perp}$ in the L-PDFT eigenstate basis. The matrices $\vb{\Theta}^{MN}$ and $\tilde{\vb{\Theta}}^{MN}$ are related to one another by the transformation matrix $\vb{U}$ that rotates the SA-CASSCF states into the L-PDFT states.
\begin{equation}
	\tilde{\vb{\Theta}}^{MN} = \transpose{\vb{U}}\vb{\Theta}^{MN}{\vb{U}}
\end{equation}
The matrix $\vb{U}$ is determined by diagonalizing the projected L-PDFT Hamiltonian in the SA-CASSCF eigenstate basis (\cref{eq:lpdft-eigenvalue}). Hence, we have that
\begin{equation}
	\pdv{E^\mathrm{SA}}{P^\Lambda_M}{P^\Gamma_N} = \frac{2}{n^\mathrm{SA}}\mel{M}{\pqty{\delta_{\Gamma \Lambda}\hat{H}^\mathrm{el}-E^\mathrm{CAS}_I U^I_\Gamma U^I_\Lambda}}{N}
\end{equation}
where $U^I_\Gamma$ are elements of the matrix $\vb{U}$. Note that $E^\mathrm{CAS}_I U^I_\Gamma U^I_\Lambda$ contains off-diagonal elements that most implementations of the SA-CASSCF Hessian matrix, which usually presume evaluation in the SA-CASSCF eigenstate basis, would omit.

\section{On-Top Gradient \label{appendix:translated-gradient}}

The first derivative of the on-top kernel ($\xck$) with respect to the density variables $\vrho$ ($\vot$) is obtained using the chain-rule
\begin{equation}
	\vot = \vxc\cdot\jacobian^{\vtrho}_{\vrho}
\end{equation}
where $\vxc$ is described by 
\begin{equation}
	\label{eq:vxc-def}
	\vxc = \nabla_{\vtrho}\xck = \begin{bmatrix}
		\pdv{\xck}{\rhoa} & \pdv{\xck}{\rhob} & \pdv{\xck}{\sigmaaa} & \pdv{\xck}{\sigmaab} & \pdv{\xck}{\sigmabb}
	\end{bmatrix}
\end{equation}
\begin{equation}
    \label{eq:formal-translated-jacobian}
	\jacobian^{\vtrho}_{\vrho} = \begin{bmatrix}
		\pdv{\rhoa}{\rho}  & \pdv{\rhoa}{\Pi}  & \pdv{\rhoa}{\rhop}  & \pdv{\rhoa}{\Pip}  \\
		\pdv{\rhob}{\rho}  & \pdv{\rhob}{\Pi}  & \pdv{\rhob}{\rhop}  & \pdv{\rhob}{\Pip}  \\
		\pdv{\sigmaaa}{\rho} & \pdv{\sigmaaa}{\Pi} & \pdv{\sigmaaa}{\rhop} & \pdv{\sigmaaa}{\Pip} \\
		\pdv{\sigmaab}{\rho} & \pdv{\sigmaab}{\Pi} & \pdv{\sigmaab}{\rhop} & \pdv{\sigmaab}{\Pip} \\ 
    \pdv{\sigmabb}{\rho} & \pdv{\sigmabb}{\Pi} & \pdv{\sigmabb}{\rhop} & \pdv{\sigmabb}{\Pip}
	\end{bmatrix}
\end{equation}
and $\jacobian^{\vtrho}_{\vrho}$ is the Jacobian matrix for the translation scheme (which has been derived previously) \cite{SandAnalytic2018,ScottAnalytic2020}.

\section{On-Top Hessian \label{appendix:translated-hessian}}

The Hessian of the on-top kernel, $\fot$, is generated from $\fxc$ via the nonlinear change of variables induced by the translation scheme:
\begin{equation}
	\fot = \transpose{\pqty{\jacobian^{\vtrho}_{\vrho}}}\cdot \fxc \cdot \jacobian^{\vtrho}_{\vrho} + \vxc\cdot \hess^{\vtrho}_{\vrho}
\end{equation}
\begin{equation}
	\fxc = \begin{bmatrix}
		\pdv[2]{\xck}{\rhoa} \\
		\pdv{\xck}{\rhoa}{\rhob}   & \pdv[2]{\xck}{\rhob} \\
    \pdv{\xck}{\rhoa}{\sigmaaa} & \pdv{\xck}{\rhob}{\sigmaaa} & \pdv[2]{\xck}{\sigmaaa} \\ 
    \pdv{\xck}{\rhoa}{\sigmaab} & \pdv{\xck}{\rhob}{\sigmaab} & \pdv{\xck}{\sigmaaa}{\sigmaab} & \pdv[2]{\xck}{\sigmaab} \\
    \pdv{\xck}{\rhoa}{\sigmabb} & \pdv{\xck}{\rhob}{\sigmabb} & \pdv{\xck}{\sigmaaa}{\sigmabb} & \pdv{\xck}{\sigmaab}{\sigmabb} & \pdv[2]{\xck}{\sigmabb}
	\end{bmatrix}
\end{equation}
where $\vxc$ is the first derivative of $\xck$ with respect to $\vtrho$ (\cref{eq:vxc-def}), $\fxc$ is defined as the symmetric matrix (here we show only the lower triangle) of the Hessian of $\xck$ with respect to the translated effective spin-density variables and their gradients, $\jacobian^{\vtrho}_{\vrho}$ is the translation Jacobian (\cref{eq:formal-translated-jacobian})\cite{SandAnalytic2018,ScottAnalytic2020}, and $\hess^{\vtrho}_{\vrho}$ is the Hessian of the translation. Both $\vxc$ and $\fxc$ are evaluated using standard KS density functional theory techniques. Specifically, in the \PySCF implementation, they are evaluated using \libxc. Throughout the rest of this section, we will only show the lower triangular portion of all Hessian matrices since they are symmetric.

Due to the complexity of these equations, we will instead derive the necessary equations through a serious of transformation with much more manageable Jacobians and Hessians. We will denote each set of coordinates (except the first and last corresponding to $\vtrho$ and $\vrho$) as $(a,b)$ which corresponds to the variables $a$, $b$, $\sigma_{aa}$, $\sigma_{ab}$, $\sigma_{bb}$. Correspondingly, we have that $\sigma$ denotes the inner product between the gradient of the two variables as
\begin{subequations}
\begin{align}
    \sigma_{aa} &= a^\prime \cdot a^\prime \\ 
    \sigma_{ab} &= a^\prime \cdot b^\prime \\ 
    \sigma_{bb} &= b^\prime \cdot b^\prime
\end{align}
\end{subequations}
Furthermore, we have that the gradient and Hessian of $\xck$ with respect to these variables are denoted as $\vb{v}^{ab}$ and $\vb{f}^{ab}$ respectively.
\begin{equation}
\begin{split}
  \vb{v}^{ab} &= \begin{bmatrix} \xck_a & \xck_b & \xck_{\sigma_{aa}} & \xck_{\sigma_{ab}} & \xck_{\sigma_{bb}} \end{bmatrix}\\ 
  &= \begin{bmatrix} \pdv{\xck}{a} & \pdv{\xck}{b} & \pdv{\xck}{\sigma_{aa}} & \pdv{\xck}{\sigma_{ab}} & \pdv{\xck}{\sigma_{bb}}\end{bmatrix}
\end{split}
\end{equation}
\begin{equation}
  \vb{f}^{ab}  = \begin{bmatrix}
    \pdv[2]{\xck}{a} \\ 
    \pdv{\xck}{a}{b} & \pdv[2]{\xck}{b} \\ 
    \pdv{\xck}{a}{\sigma_{aa}} & \pdv{\xck}{b}{\sigma_{aa}} & \pdv[2]{\xck}{\sigma_{aa}}\\ 
  \pdv{\xck}{a}{\sigma_{ab}} & \pdv{\xck}{b}{\sigma_{ab}} & \pdv{\xck}{\sigma_{aa}}{\sigma_{ab}} & \pdv[2]{\xck}{\sigma_{ab}} \\ 
    \pdv{\xck}{a}{\sigma_{bb}} & \pdv{\xck}{b}{\sigma_{bb}} & \pdv{\xck}{\sigma_{aa}}{\sigma_{bb}} & \pdv{\xck}{\sigma_{ab}}{\sigma_{bb}} & \pdv[2]{\xck}{\sigma_{bb}}
  \end{bmatrix}
\end{equation}
Furthermore, for the transformation of $(a,b)$ to $(c,d)$, we will let $\jacobian^{ab}_{cd}$ and $\hess^{ab}_{cd}$ be the corresponding Jacobian and Hessian of the transformation respectively.

The first transformation step involves going from spin-separated electron density and its derivatives to charge density ($\rho$) and spin density ($m$) and their derivatives. In this sense, $(a,b) = (\rho,m)$ so that we are translating as follows: 
\begin{equation}
  \begin{bmatrix}
    \rho \\ 
    m \\ 
    \sigma_{\rho \rho } \\ 
    \sigma_{\rho m } \\ 
    \sigma_{m m }
  \end{bmatrix} \to \begin{bmatrix}
    \rhoa \\ 
    \rhob \\ 
    \sigmaaa \\ 
    \sigmaab \\ 
    \sigmabb 
  \end{bmatrix}
\end{equation}
The coordinates are related by the following linear transformation:
\begin{equation}
    \begin{bmatrix}
        \rhoa \\ 
        \rhob \\ 
        \sigmaaa \\ 
        \sigmaab \\ 
        \sigmabb
    \end{bmatrix} = \begin{bmatrix}
        \frac{1}{2}\pqty{\rho + m} \\
        \frac{1}{2}\pqty{\rho - m} \\
        \frac{1}{4}\pqty{\sigma_{\rho\rho} + \sigma_{mm}} + \frac{1}{2}\sigma_{\rho m} \\
        \frac{1}{4}\pqty{\sigma_{\rho\rho} - \sigma_{mm}} \\
        \frac{1}{4}\pqty{\sigma_{\rho\rho} + \sigma_{mm}} - \frac{1}{2}\sigma_{\rho m}
    \end{bmatrix}
\end{equation}

Since this is a strictly linear transformation, we can see that $\fxc$ can be related to the Hessian of $\xck$ with respect to $(\rho, m)$ and their gradients as 
\begin{equation}
  \vb{f}^{\rho m} = \transpose{\pqty{\jacobian^{\uparrow\downarrow}_{\rho m}}}\cdot \fxc \cdot \jacobian^{\uparrow\downarrow}_{\rho m}
\end{equation}
\begin{equation}
  \jacobian^{\uparrow\downarrow}_{\rho m} = \frac{1}{4}\begin{bmatrix}
    2 & 2 & 0 & 0 & 0 \\
    2 & -2 & 0 & 0 & 0 \\ 
    0 & 0 & 1 & 2 & 1 \\
    0 & 0 & 1 & 0 & -1 \\ 
    0 & 0 & 1 & -2 & 1
  \end{bmatrix}
\end{equation}

Most subsequent intermediate translation steps to the coordinates $\rho$ and $\Pi$ will differ depending on whether the functional is translated or fully-translated. We first start with the simpler translated case and then go to the fully-translated case. Generally speaking though, we will undergo the following change of variables: 
\begin{enumerate}
  \item $(\rho,m)$ to $(\rho,\zeta)$ .
  \item $(\rho,\zeta)$ to $(\rho, R)$. 
  \item $(\rho, R)$ to $(\rho, \Pi)$.
  \item $(\rho, \Pi)$ to $\vrho$.
\end{enumerate}
The Hessians are related to one another by 
\begin{equation}
  \vb{f}^{\rho\zeta} = \transpose{\pqty{\jacobian^{\rho m}_{\rho \zeta}}} \cdot \vb{f}^{\rho m} \cdot \jacobian^{\rho m}_{\rho \zeta} + \vb{v}^{\rho m}\cdot \hess^{\rho m}_{\rho \zeta} 
\end{equation}
\begin{equation}
  \vb{f}^{\rho R} = \transpose{\pqty{\jacobian^{\rho \zeta}_{\rho R}}} \cdot \vb{f}^{\rho \zeta} \cdot \jacobian^{\rho \zeta}_{\rho R} + \vb{v}^{\rho \zeta}\cdot \hess^{\rho \zeta}_{\rho R} 
\end{equation}
\begin{equation}
  \vb{f}^{\rho \Pi} = \transpose{\pqty{\jacobian^{\rho R}_{\rho \Pi}}} \cdot \vb{f}^{\rho R} \cdot \jacobian^{\rho R}_{\rho \Pi} + \vb{v}^{\rho R}\cdot \hess^{\rho R}_{\rho \Pi} 
\end{equation}
\begin{equation}
    \fot = \transpose{\pqty{\jacobian^{\rho \Pi}_{\vrho}}} \cdot \vb{f}^{\rho \Pi} \cdot \jacobian^{\rho \Pi}_{\vrho} + \vb{v}^{\rho \Pi}\cdot \hess^{\rho \Pi}_{\vrho} 
\end{equation}
and the gradients are related by
\begin{equation}
  \vb{v}^{\rho m} = \vxc \cdot \jacobian^{\uparrow\downarrow}_{\rho m}
\end{equation}
\begin{equation}
  \vb{v}^{\rho\zeta} = \vb{v}^{\rho m} \cdot \jacobian^{\rho m}_{\rho \zeta}
\end{equation}
\begin{equation}
  \vb{v}^{\rho R} = \vb{v}^{\rho \zeta} \cdot \jacobian^{\rho \zeta}_{\rho R}
\end{equation}
\begin{equation}
  \vb{v}^{\rho \Pi} = \vb{v}^{\rho \Pi} \cdot \jacobian^{\rho R}_{\rho \Pi}
\end{equation}

All Jacobians and Hessians used for the translated functionals will be prefixed with a `t' (for example, $\prescript{\mathrm{t}}{}{\jacobian}$ and $\prescript{\mathrm{t}}{}{\hess}$), and Jacobians and Hessians used for fully-translated functionals will be prefixed with an `ft' (for example, $\prescript{\mathrm{ft}}{}{\jacobian}$ and $\prescript{\mathrm{ft}}{}{\hess}$). Once we arrive at the coordinates of $(\rho, \Pi)$, we can then change the variables to $\vrho$, which will be the same for translated and fully-translated functionals.

\subsection{Translated On-Top Hessian}

We now go to the variables $(\rho, \zeta)$ by 
\begin{equation}
    \label{eq:translated-rho_m-to-rho_zeta}
    \begin{bmatrix}
        \rho \\ 
        m \\ 
        \sigma_{\rho \rho} \\ 
        \sigma_{\rho m} \\ 
        \sigma_{mm}
    \end{bmatrix} = \begin{bmatrix}
        \rho \\ 
        \rho \zeta \\ 
        \sigma_{\rho \rho} \\ 
        \zeta \sigma_{\rho \rho} \\ 
        \zeta^2 \sigma_{\rho \rho}
    \end{bmatrix}
\end{equation}
Then 
\begin{equation}
  \prescript{\mathrm{t}}{}{\jacobian}^{\rho m}_{\rho \zeta} = \begin{bmatrix}
    1 & 0 & 0 & 0 & 0 \\ 
    \zeta & \rho & 0 & 0 & 0 \\ 
    0 & 0 & 1 & 0 & 0 \\ 
    0 & \sigma_{\rho\rho} & \zeta & 0 & 0 \\ 
    0 & 2\zeta\sigma_{\rho\rho} & \zeta^2 & 0 & 0
  \end{bmatrix}
\end{equation}
\begin{equation}
  \vb{v}^{\rho m} \cdot \prescript{\mathrm{t}}{}{\hess}^{\rho m}_{\rho \zeta} = \begin{bmatrix}
    0 \\ 
    \otk_m & 2\otk_{\sigma_{mm}}\sigma_{\rho\rho} \\ 
    0 & \otk_{\sigma_{\rho m}} + 2\otk_{\sigma_{mm}}\zeta & 0 \\ 
    0 & 0 & 0 & 0 \\ 
    0 & 0 & 0 & 0 & 0
  \end{bmatrix}
\end{equation}

Our next transformation is to $(\rho, R)$ using
\begin{equation}
    \label{eq:translated-rho_zeta-to-rho_R}
    \begin{bmatrix}
        \rho \\ 
        \zeta \\ 
        \sigma_{\rho \rho} \\ 
        \sigma_{\rho \zeta} \\ 
        \sigma_{\zeta\zeta}
    \end{bmatrix} = \begin{bmatrix}
        \rho \\ 
        f(R) \\ 
        \sigma_{\rho \rho} \\
        0 \\ 
        0
    \end{bmatrix}
\end{equation}
Note that in the translated case, there is no dependence on $\zeta^\prime$; therefore, the $\sigma_{\rho\zeta}$ and $\sigma_{\zeta\zeta}$ components do not contribute. Here, we are treating $f(R) = f$ and a general function of $R$ where $f^\prime = \dv{f}{R}$. Our Jacobian and Hessian for this step are 
\begin{equation}
  \prescript{\mathrm{t}}{}{\jacobian}^{\rho \zeta}_{\rho R} = \begin{bmatrix}
    1 & 0 & 0 & 0 & 0 \\ 
    0 & f^\prime & 0 & 0 & 0 \\ 
    0 & 0 & 1 & 0 & 0 \\ 
    0 & 0 & 0 & 0 & 0 \\ 
    0 & 0 & 0 & 0 & 0
  \end{bmatrix}
\end{equation}
\begin{equation}
  \vb{v}^{\rho \zeta} \cdot \prescript{\mathrm{t}}{}{\hess}^{\rho \zeta}_{\rho R } = \begin{bmatrix}
    0 \\ 
    0 & \otk_\zeta f^{\prime\prime} \\ 
    0 & 0 & 0 \\ 
    0 & 0 & 0 & 0 \\ 
    0 & 0 & 0 & 0 & 0
  \end{bmatrix}
\end{equation}
where $f^{\prime\prime}$ is the second derivative of $\zeta$ with respect to $R$.

Next we go to the $(\rho, \Pi)$ variables by the following transformation:
\begin{equation}
    \label{eq:translated-rho_R-to-rho_Pi}
    \begin{bmatrix}
        \rho \\ 
        R \\ 
        \sigma_{\rho\rho} \\ 
        \sigma_{\rho R} \\ 
        \sigma_{RR}
    \end{bmatrix} = \begin{bmatrix}
        \rho \\ 
        \frac{4\Pi}{\rho^2} \\ 
        \sigma_{\rho\rho} \\
        0 \\ 
        0
    \end{bmatrix}
\end{equation}
Again, we can omit the $\sigma_{\rho R}$ and $\sigma_{RR}$ variables since translated functionals do not depend on $R^\prime$. This transformation results in 
\begin{equation}
  \prescript{\mathrm{t}}{}{\jacobian}^{\rho R}_{\rho \Pi} = \frac{4}{\rho^2}\begin{bmatrix}
    1 & 0 & 0 & 0 & 0 \\ 
    \frac{-2\Pi}{\rho} & 1 & 0 & 0 & 0 \\ 
    0 & 0 & 1 & 0 & 0 \\ 
    0 & 0 & 0 & 0 & 0 \\ 
    0 & 0 & 0 & 0 & 0
  \end{bmatrix}
\end{equation}
\begin{equation}
  \vb{v}^{\rho R} \cdot \prescript{\mathrm{t}}{}{\hess}^{\rho R}_{\rho \Pi} = \frac{2\otk_R}{\rho^2}\begin{bmatrix}
    3R \\ 
    \frac{-4}{\rho} & 0 \\ 
    0 & 0 & 0 \\ 
    0 & 0 & 0 & 0 \\ 
    0 & 0 & 0 & 0 & 0
  \end{bmatrix}
\end{equation}

\subsection{Fully-Translated On-Top Hessian}

For the fully-translated case, going to the $(\rho, m)$ variables modifies $\sigma_{\rho m}$ and $\sigma_{mm}$ in \cref{eq:translated-rho_m-to-rho_zeta} such that 
\begin{equation}
    \begin{bmatrix}
        \rho \\ 
        m \\ 
        \sigma_{\rho \rho} \\ 
        \sigma_{\rho m} \\ 
        \sigma_{mm}
    \end{bmatrix} = \begin{bmatrix}
        \rho \\ 
        \rho \zeta \\ 
        \sigma_{\rho \rho} \\ 
        \zeta\sigma_{\rho\rho} + \rho\sigma_{\rho\zeta} \\ 
        \zeta^2\sigma_{\rho\rho} + 2\rho\zeta\sigma_{\rho\zeta} + \rho^2\sigma_{\zeta\zeta}
    \end{bmatrix}
\end{equation}
The fully-translated Jacobian and Hessian for this translation step are slight modifications of the translated matrices.
\begin{equation}
  \prescript{\mathrm{ft}}{}{\jacobian}^{\rho m}_{\rho \zeta} = \prescript{\mathrm{t}}{}{\jacobian}^{\rho m}_{\rho \zeta} + \tilde{\jacobian}^{\rho m}_{\rho\zeta} 
\end{equation}
\begin{equation}
  \tilde{\jacobian}^{\rho m}_{\rho \zeta} = \begin{bmatrix}
    0 & 0 & 0 & 0 & 0 \\ 
    0 & 0 & 0 & 0 & 0 \\ 
    0 & 0 & 0 & 0 & 0 \\ 
    \sigma_{\rho\zeta} & 0 & 0 & \rho & 0 \\ 
    2\pqty{\rho\sigma_{\zeta\zeta} + \zeta\sigma_{\rho\zeta}} & 2\rho\sigma_{\rho\zeta} & 0 & 2\rho\zeta & \rho^2
  \end{bmatrix}
\end{equation}
\begin{equation}
  \vb{v}^{\rho m}\cdot \prescript{\mathrm{ft}}{}{\hess}^{\rho m}_{\rho\zeta} = \vb{v}^{\rho m} \cdot \prescript{\mathrm{t}}{}{\hess}^{\rho m}_{\rho\zeta} + \vb{v}^{\rho m} \cdot \tilde{\hess}^{\rho m}_{\rho\zeta}
\end{equation}
\begin{equation}
  \vb{v}^{\rho m}\cdot\tilde{\hess}^{\rho m}_{\rho\zeta} = \begin{bmatrix}
    2\otk_{\sigma_{mm}}\sigma_{\zeta\zeta} \\ 
    2\otk_{\sigma_{mm}}\sigma_{\rho\zeta} & 0 \\ 
    0 & 0 & 0 \\ 
    \otk_{\sigma_{\rho m}} + 2\otk_{\sigma_{mm}}\zeta & 2\otk_{\sigma_{mm}}\rho & 0 & 0 \\ 
    2\otk_{\sigma_{mm}}\rho & 0 & 0 & 0 & 0
  \end{bmatrix}
\end{equation}

For the next transformation step to $(\rho, R)$, we must include the transformations for $\sigma_{\rho\zeta}$ and $\sigma_{\zeta\zeta}$ so that \cref{eq:translated-rho_zeta-to-rho_R} is modified to be
\begin{equation}
    \begin{bmatrix}
        \rho \\ 
        \zeta \\ 
        \sigma_{\rho \rho} \\ 
        \sigma_{\rho \zeta} \\ 
        \sigma_{\zeta\zeta}
    \end{bmatrix} = \begin{bmatrix}
        \rho \\ 
        f(R) \\ 
        \sigma_{\rho \rho} \\
        f^\prime \sigma_{\rho R} \\ 
        \pqty{f^\prime}^2\sigma_{RR} 
    \end{bmatrix}
\end{equation}
This leads to the modified Jacobian and Hessian as 
\begin{equation}
  \prescript{\mathrm{ft}}{}{\jacobian}^{\rho\zeta}_{\rho R} = \prescript{\mathrm{t}}{}{\jacobian}^{\rho\zeta}_{\rho R} + \tilde{\jacobian}^{\rho\zeta}_{\rho R}
\end{equation}
\begin{equation}
  \tilde{\jacobian}^{\rho\zeta}_{\rho R} = \begin{bmatrix}
    0 & 0 & 0 & 0 & 0 \\ 
    0 & 0 & 0 & 0 & 0 \\ 
    0 & 0 & 0 & 0 & 0 \\ 
    0 & \sigma_{\rho R}f^{\prime\prime} & 0 & f^{\prime} & 0 \\ 
    0 & 2\sigma_{RR}f^{\prime\prime}f^{\prime} & 0 & 0 & \pqty{f^\prime}^2
  \end{bmatrix}
\end{equation}
\begin{equation}
  \vb{v}^{\rho\zeta}\cdot \prescript{\mathrm{ft}}{}{\hess}^{\rho\zeta}_{\rho R} = \vb{v}^{\rho \zeta}\cdot \prescript{\mathrm{t}}{}{\hess}^{\rho \zeta}_{\rho R} + \vb{v}^{\rho \zeta}\cdot \tilde{\hess}^{\rho\zeta}_{\rho R}
\end{equation}
\begin{equation}
  \vb{v}^{\rho \zeta}\cdot \tilde{\hess}^{\rho\zeta}_{\rho R} = \otk_{\sigma_{\rho\zeta}} \hess^{\sigma_{\rho\zeta}}_{\rho R} + \otk_{\sigma_{\zeta\zeta}} \hess^{\sigma_{\zeta\zeta}}_{\rho R} 
\end{equation}
where $\hess^{\sigma_{\rho \zeta}}_{\rho R}$ and $\hess^{\sigma_{\zeta\zeta}}_{\rho R}$ are the Hessians of $\sigma_{\rho\zeta}$ and $\sigma_{\zeta\zeta}$ with respect to the $(\rho, R)$ variables.
\begin{equation}
  \hess^{\sigma_{\rho\zeta}}_{\rho R} = \begin{bmatrix}
    0 \\ 
    0 & \sigma_{\rho R}f^{\prime\prime\prime} \\ 
    0 & 0 & 0 \\ 
    0 & f^{\prime\prime} & 0 & 0 \\ 
    0 & 0 & 0 & 0 & 0
  \end{bmatrix} 
\end{equation}
\begin{align}
  \hess^{\sigma_{\zeta\zeta}}_{\rho R} = \begin{bmatrix}
    0 \\ 
    0 & 2\sigma_{RR}\pqty{f^{\prime\prime\prime}f^{\prime} + \pqty{f^{\prime\prime}}^2} \\ 
    0 & 0 & 0 \\ 
    0 & 0 & 0 & 0 \\ 
    0 & 2f^{\prime\prime}f^{\prime} & 0 & 0 & 0
  \end{bmatrix} 
\end{align}

The final transformation we must consider separately for the fully-translated functionals is to the $(\rho, \Pi)$ coordinate. Here, we must include the transformation of the $\sigma_{\rho R}$ and $\sigma_{R R}$, which are not included in the translated case. The modified form of \cref{eq:translated-rho_R-to-rho_Pi} for the fully-translated case is thus
\begin{equation}
    \begin{bmatrix}
        \rho \\ 
        R \\ 
        \sigma_{\rho\rho} \\ 
        \sigma_{\rho R} \\ 
        \sigma_{RR}
    \end{bmatrix} = \begin{bmatrix}
        \rho \\ 
        \frac{4\Pi}{\rho^2} \\ 
        \sigma_{\rho\rho} \\
        \frac{4}{\rho^2}\pqty{\sigma_{\rho \Pi} - \frac{2\Pi}{\rho}\sigma_{\rho\rho}} \\ 
        \frac{16}{\rho^4}\pqty{\frac{4\Pi^2}{\rho^2}\sigma_{\rho\rho} - \frac{4\Pi}{\rho}\sigma_{\rho\Pi} + \sigma_{\Pi\Pi}}
    \end{bmatrix}
\end{equation}
The fully-translated Jacobian is given by 
\begin{equation}
  \prescript{\mathrm{ft}}{}{\jacobian}^{\rho R}_{\rho \Pi} = \prescript{\mathrm{t}}{}{\jacobian}^{\rho R}_{\rho \Pi} + \tilde{\jacobian}^{\rho R}_{\rho \Pi} 
\end{equation}
\begin{widetext}
\begin{equation}
  \tilde{\jacobian}^{\rho R}_{\rho \Pi} = \frac{4}{\rho^2}\begin{bmatrix}
    0 & 0 & 0 & 0 & 0 \\ 
    0 & 0 & 0 & 0 & 0 \\ 
    0 & 0 & 0 & 0 & 0 \\ 
    \frac{2}{\rho}\pqty{\frac{3\Pi\sigma_{\rho\rho}}{\rho} - \sigma_{\rho\Pi}} & \frac{-2\sigma_{\rho\rho}}{\rho} & \frac{-2\Pi}{\rho} & 1 & 0 \\ 
    \frac{16}{\rho^3}\pqty{\frac{5\Pi\sigma_{\rho\Pi}}{\rho} - \frac{6\Pi^2\sigma_{\rho\rho}}{\rho^2} - \sigma_{\Pi\Pi}} & \frac{16}{\rho^3}\pqty{\frac{2\Pi\sigma_{\rho\rho}}{\rho} - \sigma_{\rho\Pi}} & \frac{16\Pi^2}{\rho^4} & \frac{-16\Pi}{\rho^3} & \frac{4}{\rho^2}
  \end{bmatrix} 
\end{equation}
\end{widetext}
And the fully-translated Hessian term is given by 
\begin{equation}
  \vb{v}^{\rho R} \cdot \prescript{\mathrm{ft}}{}{\hess}^{\rho R}_{\rho \Pi} = \vb{v}^{\rho R} \cdot \prescript{\mathrm{t}}{}{\hess}^{\rho R}_{\rho \Pi} + \vb{v}^{\rho R} \cdot \tilde{\hess}^{\rho R}_{\rho \Pi} 
\end{equation}
\begin{equation}
  \vb{v}^{\rho R}\cdot \tilde{\hess}^{\rho R}_{\rho \Pi} = \otk_{\sigma_{\rho R}} \hess^{\sigma_{\rho R}}_{\rho\Pi} + \otk_{\sigma_{R R}}\hess^{\sigma_{RR}}_{\rho\Pi}
\end{equation}
with $\hess^{\sigma_{\rho R}}_{\rho\Pi}$ and $\hess^{\sigma_{RR}}_{\rho \Pi}$ the Hessian of $\sigma_{\rho R}$ and $\sigma_{RR}$ with respect to the $(\rho, \Pi)$ variables given by
\begin{equation}
  \hess^{\sigma_{\rho R}}_{\rho \Pi} = \begin{bmatrix}
    \frac{24}{\rho^3}\pqty{\sigma_{\rho\Pi} - R\sigma_{\rho\rho}} \\ 
    \frac{24\sigma_{\rho\rho}}{\rho^4} & 0 \\ 
    \frac{6R}{\rho^2} & \frac{-8}{\rho^3} & 0 \\ 
    \frac{-8}{\rho^3} & 0 & 0 & 0 \\ 
    0 & 0 & 0 & 0 & 0
  \end{bmatrix} 
\end{equation}
\begin{equation}
  \hess^{\sigma_{RR}}_{\rho \Pi} = \frac{8}{\rho^3}\begin{bmatrix}
    \frac{21R^2\sigma_{\rho\rho}}{\rho} - \frac{60R\sigma_{\rho\Pi}}{\rho^2} + \frac{40\sigma_{\Pi\Pi}}{\rho^3} \\ 
    \frac{8}{\rho^2}\pqty{\frac{5\sigma_{\rho\Pi}}{\rho} - 3R\sigma_{\rho\rho} } & \frac{16\sigma_{\rho\rho}}{\rho^3} \\ 
    -3R^2 & \frac{4R}{\rho} & 0 \\ 
    \frac{10R}{\rho} & \frac{-8}{\rho^2} & 0 & 0 \\ 
    \frac{-8}{\rho^2} & 0 & 0 & 0 & 0
  \end{bmatrix} 
\end{equation}

\subsection{Unpacking the Sigma Vector}

At this point in both the translated and fully-translated cases, we have arrived at the Hessian of $\xck$ with respect to $\rho$, $\Pi$, $\sigma_{\rho\rho}$, $\sigma_{\rho\Pi}$, and $\sigma_{\Pi\Pi}$. It is fairly easy to transform to the canonical variables of $\vrho$ by noting that 
\begin{equation}
  \sigma_{\rho\rho} = \rhop \cdot \rhop
\end{equation}
\begin{equation}
  \sigma_{\rho\Pi} = \rhop \cdot \Pip
\end{equation}
\begin{equation}
  \sigma_{\Pi\Pi} = \Pip \cdot \Pip
\end{equation}
Hence, we have that 
\begin{equation}
  \fot = \transpose{\pqty{\jacobian^{\rho\Pi}_{\vrho}}}\cdot \vb{f}^{\rho\Pi} \cdot \jacobian^{\rho \Pi}_{\vrho} + \vb{v}^{\rho\Pi} \cdot \hess^{\rho \Pi}_{\vrho}
\end{equation}
\begin{equation}
  \jacobian^{\rho\Pi}_{\vrho} = \begin{bmatrix}
    1 & 0 & 0 & 0 & 0 \\ 
    0 & 1 & 0 & 0 & 0 \\ 
    0 & 0 & 2\rhop & \Pip & 0 \\ 
    0 & 0 & 0 & \rhop & 2\Pip
  \end{bmatrix}
\end{equation}
\begin{equation}
  \vb{v}^{\rho\Pi}\cdot \hess^{\rho\Pi}_{\vrho} = \begin{bmatrix}
    0 \\  
    0 & 0  \\ 
    0 & 0 & 2\otk_{\sigma_{\rho\rho}} \\ 
    0 & 0 & \otk_{\sigma_{\rho\Pi}} & 2\otk_{\sigma_{\Pi\Pi}}
  \end{bmatrix}
\end{equation} 

\bibliography{ms}

\section*{TOC Graphic}
\includegraphics{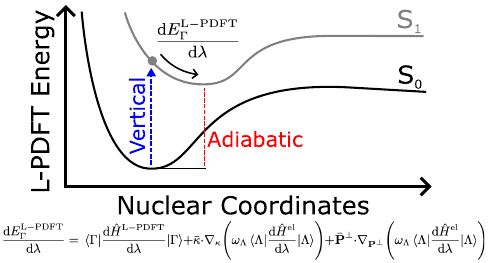}

\end{document}


\title{Supplementary Material: Analytic Nuclear Gradients for Complete Active Space Linearized Pair-Density Functional Theory}

\author{Matthew R. Hennefarth}
\author{Matthew R. Hermes}
\affiliation{Department of Chemistry and Chicago Center for Theoretical Chemistry, University of Chicago, Chicago, IL 60637, USA}

\author{Donald G. Truhlar}
 \email[corresponding author: ]{truhlar@umn.edu}
\affiliation{Department of Chemistry, Chemical Theory Center, and Minnesota Supercomputing Institute, University of Minnesota, Minneapolis, MN 55455-0431, USA}

\author{Laura Gagliardi}
 \email[corresponding author: ]{lgagliardi@uchicago.edu}
\affiliation{Department of Chemistry, Pritzker School of Molecular Engineering, The James Franck Institute, and Chicago Center for Theoretical Chemistry, University of Chicago, Chicago, IL 60637, USA}
\affiliation{Argonne National Laboratory, 9700 S. Cass Avenue, Lemont, IL 60439, USA}

\date{March 14, 2024}

\maketitle

\tableofcontents

\section{Supplemental Figures}
\begin{figure}[H]
    \includegraphics[width=\columnwidth]{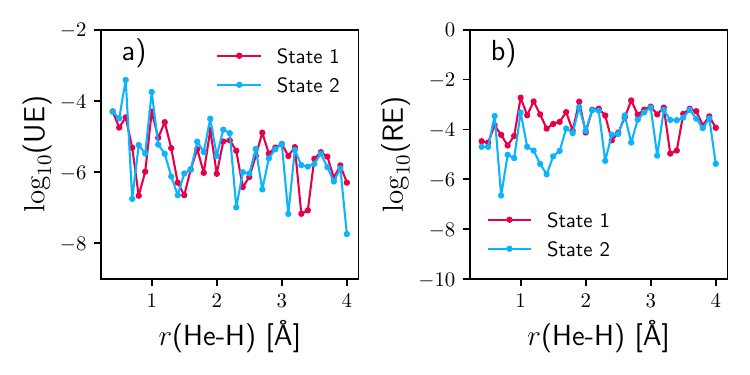}
    \caption{\label{sfig:heh-err-tpbe} Log unsigned (a) and relative (b) error for \ce{HeH+} computed with the tPBE functional at various He-H distances.}
\end{figure}

\begin{figure}[H]
    \includegraphics[width=\columnwidth]{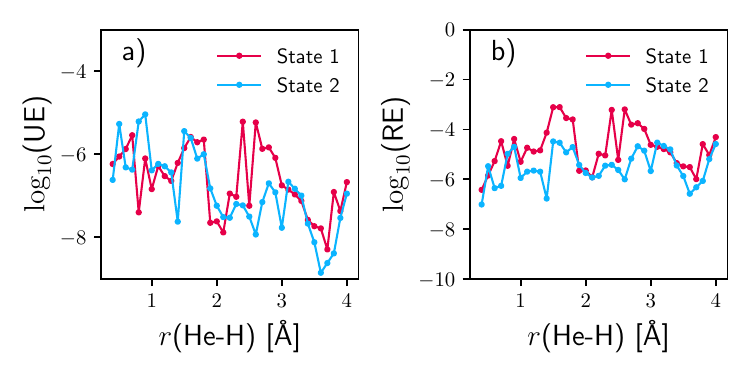}
    \caption{\label{sfig:heh-err-ftlda} Log unsigned (a) and relative (b) error for \ce{HeH+} computed with the ftSVWN3 functional at various He-H distances.}
\end{figure}

\begin{figure}[H]
    \includegraphics[width=\columnwidth]{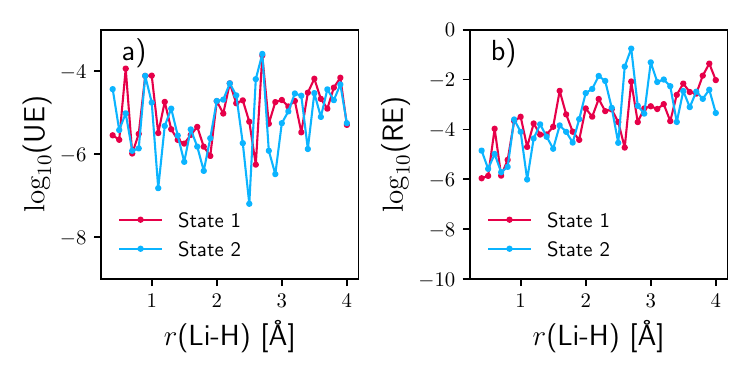}
    \caption{\label{sfig:lih-err-tpbe} Log unsigned (a) and relative (b) error for \ce{LiH} computed with the tPBE functional at various Li-H distances.}
\end{figure}

\begin{figure}[H]
    \includegraphics[width=\columnwidth]{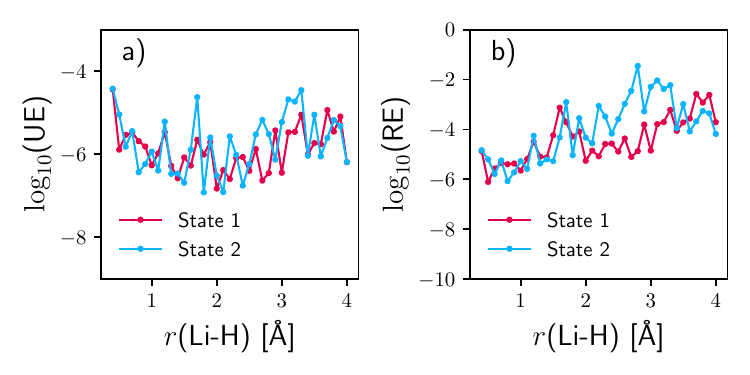}
    \caption{\label{sfig:lih-err-ftpbe} Log unsigned (a) and relative (b) error for \ce{LiH} computed with the ftPBE functional at various Li-H distances.}
\end{figure}

\section{Optimized Geometries}

\subsection{Formaldehyde}

\begin{table}[H]
\centering
\caption{L-PDFT formaldehyde equilibrium ground state geometry (in \unit{\angstrom}). State 0 energy: \qty{-114.38370839}{\hartree}. State 1 energy: \qty{-114.23739537}{\hartree}}
\begin{tabular}{l d d d}
C & 0.49586 & -0.00001 & -0.00001 \\
O & -0.71370 & 0.00001 & -0.00004 \\
H & 1.08345 & -0.00000 & 0.94756 \\
H & 1.08356 & 0.00000 & -0.94751 \\
\end{tabular}
\end{table}

\begin{table}[H]
\centering
\caption{L-PDFT formaldehyde equilibrium excited state geometry (in \unit{\angstrom}). State 0 energy: \qty{-114.35572479}{\hartree}. State 1 energy: \qty{-114.25098675}{\hartree}}
\begin{tabular}{l d d d}
C & 0.50356 & -0.02576 & -0.00008 \\
O & -0.64094 & 0.64726 & -0.00003 \\
H & 1.06843 & 0.01424 & 0.94321 \\
H & 1.06836 & 0.01426 & -0.94341 \\
\end{tabular}
\end{table}

\subsection{Phenol}

\begin{table}[H]
\centering
\caption{L-PDFT phenol equilibrium ground state geometry (in \unit{\angstrom}). State 0 energy: \qty{-307.06157388}{\hartree}. State 1 energy: \qty{-306.87656579}{\hartree}. State 2 energy: \qty{-306.85927128}{\hartree}.}
\begin{tabular}{l d d d}
C & 1.26641 & 0.00000 & 2.08922 \\
C & 2.48510 & 0.00000 & 1.39749 \\
C & 2.47503 & -0.00000 & -0.00203 \\
C & 1.26674 & -0.00000 & -0.70531 \\
C & 0.05247 & -0.00000 & -0.00139 \\
C & 0.05113 & -0.00000 & 1.40115 \\
O & -1.16160 & -0.00000 & -0.63308 \\
H & 3.42040 & -0.00000 & -0.55945 \\
H & 3.43496 & 0.00000 & 1.94523 \\
H & 1.25970 & 0.00000 & 3.18668 \\
H & -0.90866 & 0.00000 & 1.93095 \\
H & 1.26439 & -0.00000 & -1.80482 \\
H & -1.02640 & 0.00000 & -1.58516 \\
\end{tabular}
\end{table}

\begin{table}[H]
\centering
\caption{L-PDFT phenol equilibrium first excited state geometry (in \unit{\angstrom}). State 0 energy: \qty{-307.05461258}{\hartree}. State 1 energy: \qty{-306.88339667}{\hartree}. State 2 energy: \qty{-306.85866671}{\hartree}.}
\begin{tabular}{l d d d}
C & 1.25874 & 0.00000 & 2.13850 \\
C & 2.47518 & 0.00000 & 1.39403 \\
C & 2.50474 & 0.00000 & -0.03457 \\
C & 1.26890 & 0.00000 & -0.75821 \\
C & 0.05105 & -0.00000 & 0.00454 \\
C & 0.00947 & -0.00000 & 1.43296 \\
O & -1.13755 & -0.00000 & -0.62615 \\
H & 3.46259 & 0.00000 & -0.56442 \\
H & 3.42850 & 0.00000 & 1.93823 \\
H & 1.28146 & 0.00000 & 3.23263 \\
H & -0.96178 & -0.00000 & 1.93597 \\
H & 1.23317 & 0.00000 & -1.85461 \\
H & -0.99479 & -0.00000 & -1.57942 \\
\end{tabular}
\end{table}

\subsection{\textit{Trans}-butadiene}

\begin{table}[H]
\centering
\caption{L-PDFT \textit{trans}-butadiene equilibrium ground state (1 \textsuperscript{1}A\textsubscript{g}) geometry (in \unit{\angstrom}). 1 \textsuperscript{1}A\textsubscript{g} energy: \qty{-155.78711361}{\hartree}. 2 \textsuperscript{1}A\textsubscript{g} energy: \qty{-155.53293652}{\hartree}.}
\begin{tabular}{l d d d}
C & 1.16325 & 0.00002 & -1.43351 \\
C & 0.02896 & -0.00001 & -0.72932 \\
C & -0.02896 & 0.00001 & 0.72932 \\
C & -1.16325 & -0.00002 & 1.43351 \\
H & -0.93173 & -0.00004 & -1.25031 \\
H & 1.16211 & 0.00001 & -2.52157 \\
H & 2.13664 & 0.00006 & -0.94202 \\
H & 0.93173 & 0.00004 & 1.25031 \\
H & -1.16211 & -0.00001 & 2.52157 \\
H & -2.13664 & -0.00006 & 0.94202 \\
\end{tabular}
\end{table}

\begin{table}[H]
\centering
\caption{L-PDFT \textit{trans}-butadiene equilibrium 2 \textsuperscript{1}A\textsubscript{g} state geometry (in \unit{\angstrom}). 1 \textsuperscript{1}A\textsubscript{g} energy: \qty{-155.74867922}{\hartree}. 2 \textsuperscript{1}A\textsubscript{g} energy: \qty{-155.57485703}{\hartree}.}
\begin{tabular}{l d d d}
C & 1.26795 & 0.00003 & -1.51460 \\
C & 0.01332 & -0.00001 & -0.69913 \\
C & -0.01332 & 0.00001 & 0.69913 \\
C & -1.26795 & -0.00003 & 1.51460 \\
H & -0.93183 & -0.00004 & -1.24190 \\
H & 1.21814 & 0.00002 & -2.59769 \\
H & 2.24001 & 0.00006 & -1.03038 \\
H & 0.93183 & 0.00004 & 1.24190 \\
H & -1.21814 & -0.00002 & 2.59769 \\
H & -2.24001 & -0.00006 & 1.03038 \\
\end{tabular}
\end{table}

\subsection{Cytosine}

\begin{table}[H]
\centering
\caption{L-PDFT cytosine equilibrium ground state (1 \textsuperscript{1}A\textsuperscript{$\prime$}) geometry (in \unit{\angstrom}). 1 \textsuperscript{1}A\textsuperscript{$\prime$} energy: \qty{-394.55508466}{\hartree}. 2 \textsuperscript{1}A\textsuperscript{$\prime$} energy: \qty{-394.39281789}{\hartree}. 3 \textsuperscript{1}A\textsuperscript{$\prime$} energy: \qty{-394.36098028}{\hartree}.}
\begin{tabular}{l d d d}
C & 1.04984 & 1.19134 & -0.00416 \\
C & 1.13724 & -0.23990 & -0.00508 \\
N & 0.08654 & -1.05068 & -0.00078 \\
C & -1.17747 & -0.54080 & 0.00496 \\
N & -1.27884 & 0.89327 & 0.00595 \\
C & -0.20931 & 1.70775 & 0.00157 \\
N & 2.35051 & -0.83542 & -0.01063 \\
O & -2.21720 & -1.17514 & 0.00929 \\
H & -2.22563 & 1.25388 & 0.01024 \\
H & -0.40459 & 2.77816 & 0.00283 \\
H & 1.92547 & 1.83177 & -0.00776 \\
H & 3.20773 & -0.30568 & -0.01419 \\
H & 2.38152 & -1.84599 & -0.01115 \\
\end{tabular}
\end{table}

\begin{table}[H]
\centering
\caption{L-PDFT cytosine equilibrium 2 \textsuperscript{1}A\textsuperscript{$\prime$} state geometry (in \unit{\angstrom}). 1 \textsuperscript{1}A\textsuperscript{$\prime$} energy: \qty{-394.53676415}{\hartree}. 2 \textsuperscript{1}A\textsuperscript{$\prime$} energy: \qty{-394.4062025}{\hartree}. 3 \textsuperscript{1}A\textsuperscript{$\prime$} energy: \qty{-394.36145003}{\hartree}.}
\begin{tabular}{l d d d}
C & 1.05299 & 1.18617 & -0.00417 \\
C & 1.18415 & -0.18352 & -0.00527 \\
N & 0.07770 & -1.05043 & -0.00074 \\
C & -1.11395 & -0.49925 & 0.00470 \\
N & -1.30540 & 0.88580 & 0.00606 \\
C & -0.23454 & 1.77191 & 0.00170 \\
N & 2.36308 & -0.86413 & -0.01070 \\
O & -2.19194 & -1.17166 & 0.00918 \\
H & -2.27061 & 1.19130 & 0.01042 \\
H & -0.43974 & 2.83438 & 0.00300 \\
H & 1.92625 & 1.83475 & -0.00776 \\
H & 3.25747 & -0.39974 & -0.01445 \\
H & 2.32035 & -1.87302 & -0.01089 \\
\end{tabular}
\end{table}

\section{Absolute Gradients}

\begin{table}[H]
\centering
\caption{Analytical and extrapolated numerical gradients (hartree/bohr) for \ce{HeH+} with the tPBE functional at various He-H distances (\AA).}
\begin{ruledtabular}
\begin{tabular}{S[table-format=-1.1]S[table-format=-1.10]S[table-format=-1.10]S[table-format=-1.10]S[table-format=-1.10]}
{} & \multicolumn{2}{c}{Analytic} & \multicolumn{2}{c}{Numeric (extrapolated)} \\
{R} & {State 1} & {State 2} & {State 1} & {State 2} \\
\hline
0.4 & -1.5226469532 & -2.4864239754 & -1.5225962956 & -2.4864735687 \\
0.5 & -0.6052792212 & -1.6238002608 & -0.6052614335 & -1.6238324966 \\
0.6 & -0.2387700531 & -1.1148977564 & -0.2388045176 & -1.1152846125 \\
0.7 & -0.0805289066 & -0.7978183597 & -0.0805240792 & -0.7978185358 \\
0.8 & -0.0095877493 & -0.6013286411 & -0.0095875329 & -0.6013228965 \\
0.9 & 0.0188922265 & -0.4693833012 & 0.0188912010 & -0.4693799981 \\
1.0 & 0.0266119576 & -0.3723024562 & 0.0266617535 & -0.3721245751 \\
1.1 & 0.0249039992 & -0.2961763679 & 0.0249131805 & -0.2961822777 \\
1.2 & 0.0190563949 & -0.2344032921 & 0.0190815257 & -0.2344065600 \\
1.3 & 0.0118411795 & -0.1840489653 & 0.0118458778 & -0.1840497174 \\
1.4 & 0.0046957613 & -0.1430224131 & 0.0046962639 & -0.1430226364 \\
1.5 & -0.0013590270 & -0.1105192556 & -0.0013588033 & -0.1105201687 \\
1.6 & -0.0059058770 & -0.0852613282 & -0.0059070635 & -0.0852601593 \\
1.7 & -0.0089571752 & -0.0659621688 & -0.0089615762 & -0.0659550257 \\
1.8 & -0.0107524886 & -0.0515497486 & -0.0107534420 & -0.0515460900 \\
1.9 & -0.0116698351 & -0.0406499709 & -0.0116547122 & -0.0406814294 \\
2.0 & -0.0119471322 & -0.0326296409 & -0.0119480280 & -0.0326267700 \\
2.1 & -0.0118428070 & -0.0266248519 & -0.0118355386 & -0.0266404230 \\
2.2 & -0.0114988135 & -0.0221053895 & -0.0115066003 & -0.0220929617 \\
2.3 & -0.0110246100 & -0.0186845464 & -0.0110206445 & -0.0186844446 \\
2.4 & -0.0105060998 & -0.0159981395 & -0.0105064772 & -0.0159971518 \\
2.5 & -0.0099584394 & -0.0139260512 & -0.0099577145 & -0.0139269566 \\
2.6 & -0.0094206573 & -0.0122738431 & -0.0094178175 & -0.0122782737 \\
2.7 & -0.0088968139 & -0.0109441073 & -0.0088840092 & -0.0109437827 \\
2.8 & -0.0084046780 & -0.0098574553 & -0.0084079775 & -0.0098598755 \\
2.9 & -0.0079258988 & -0.0089654807 & -0.0079210251 & -0.0089698517 \\
3.0 & -0.0074884133 & -0.0082102022 & -0.0074945460 & -0.0082040771 \\
3.1 & -0.0070619318 & -0.0075822677 & -0.0070591077 & -0.0075822011 \\
3.2 & -0.0066757733 & -0.0070277733 & -0.0066707694 & -0.0070319594 \\
3.3 & -0.0063118533 & -0.0065522886 & -0.0063119207 & -0.0065538528 \\
3.4 & -0.0059651116 & -0.0061367304 & -0.0059651955 & -0.0061381581 \\
3.5 & -0.0056550256 & -0.0057570889 & -0.0056573897 & -0.0057553820 \\
3.6 & -0.0053571115 & -0.0054261747 & -0.0053534985 & -0.0054294760 \\
3.7 & -0.0050812577 & -0.0051242642 & -0.0050785477 & -0.0051256383 \\
3.8 & -0.0048256933 & -0.0048494428 & -0.0048249998 & -0.0048499866 \\
3.9 & -0.0045890063 & -0.0045969857 & -0.0045874733 & -0.0045982614 \\
4.0 & -0.0043661064 & -0.0043673805 & -0.0043656029 & -0.0043673986 \\
\end{tabular}
\end{ruledtabular}
\end{table}

\begin{table}[H]
\centering
\caption{Analytical and extrapolated numerical gradients (hartree/bohr) for \ce{HeH+} with the ftSVWN3 functional at various He-H distances (\AA).}
\begin{ruledtabular}
\begin{tabular}{S[table-format=-1.1]S[table-format=-1.10]S[table-format=-1.10]S[table-format=-1.10]S[table-format=-1.10]}
{} & \multicolumn{2}{c}{Analytic} & \multicolumn{2}{c}{Numeric (extrapolated)} \\
{R} & {State 1} & {State 2} & {State 1} & {State 2} \\
\hline
0.4 & -1.5575011318 & -2.4977997106 & -1.5575017147 & -2.4977999520 \\
0.5 & -0.6278903968 & -1.6202330061 & -0.6278912787 & -1.6202383686 \\
0.6 & -0.2510087310 & -1.1067325644 & -0.2510100689 & -1.1067320809 \\
0.7 & -0.0859977456 & -0.7879128402 & -0.0859948755 & -0.7879124141 \\
0.8 & -0.0115410083 & -0.5906362988 & -0.0115409685 & -0.5906301485 \\
0.9 & 0.0193002204 & -0.4602938553 & 0.0193010123 & -0.4602847105 \\
1.0 & 0.0289030357 & -0.3659431103 & 0.0289028911 & -0.3659427005 \\
1.1 & 0.0283673546 & -0.2920936262 & 0.0283668369 & -0.2920942117 \\
1.2 & 0.0231389596 & -0.2320982141 & 0.0231386620 & -0.2320987276 \\
1.3 & 0.0159052606 & -0.1827417589 & 0.0159054884 & -0.1827421239 \\
1.4 & 0.0083855801 & -0.1424703376 & 0.0083849616 & -0.1424703614 \\
1.5 & 0.0018259817 & -0.1103617009 & 0.0018245750 & -0.1103580917 \\
1.6 & -0.0033446780 & -0.0851576037 & -0.0033472805 & -0.0851551373 \\
1.7 & -0.0068352007 & -0.0659457543 & -0.0068371479 & -0.0659449699 \\
1.8 & -0.0090631571 & -0.0514231625 & -0.0090654179 & -0.0514221625 \\
1.9 & -0.0103055162 & -0.0405082627 & -0.0103054938 & -0.0405084136 \\
2.0 & -0.0108454601 & -0.0324131690 & -0.0108454843 & -0.0324132266 \\
2.1 & -0.0110041558 & -0.0262864251 & -0.0110041689 & -0.0262864558 \\
2.2 & -0.0108272105 & -0.0217316593 & -0.0108270973 & -0.0217316889 \\
2.3 & -0.0105441435 & -0.0182267056 & -0.0105442381 & -0.0182266421 \\
2.4 & -0.0101251211 & -0.0155719154 & -0.0101190280 & -0.0155719742 \\
2.5 & -0.0096905387 & -0.0134962556 & -0.0096905962 & -0.0134962240 \\
2.6 & -0.0092246731 & -0.0118805296 & -0.0092305032 & -0.0118805413 \\
2.7 & -0.0087731655 & -0.0105829351 & -0.0087745201 & -0.0105830056 \\
2.8 & -0.0083222536 & -0.0095434371 & -0.0083207884 & -0.0095436368 \\
2.9 & -0.0078835327 & -0.0086922760 & -0.0078827121 & -0.0086923973 \\
3.0 & -0.0074579973 & -0.0079893689 & -0.0074578190 & -0.0079893519 \\
3.1 & -0.0070536595 & -0.0073927613 & -0.0070535184 & -0.0073929791 \\
3.2 & -0.0066720500 & -0.0068794576 & -0.0066719423 & -0.0068796053 \\
3.3 & -0.0063097096 & -0.0064346388 & -0.0063096342 & -0.0064347398 \\
3.4 & -0.0059717059 & -0.0060401998 & -0.0059717322 & -0.0060402212 \\
3.5 & -0.0056558625 & -0.0056875921 & -0.0056558810 & -0.0056875997 \\
3.6 & -0.0053600075 & -0.0053709490 & -0.0053600239 & -0.0053709476 \\
3.7 & -0.0050851488 & -0.0050819112 & -0.0050851437 & -0.0050819136 \\
3.8 & -0.0048290266 & -0.0048178924 & -0.0048289033 & -0.0048178965 \\
3.9 & -0.0045901209 & -0.0045752691 & -0.0045901638 & -0.0045752984 \\
4.0 & -0.0043677521 & -0.0043506967 & -0.0043679656 & -0.0043505841 \\
\end{tabular}
\end{ruledtabular}
\end{table}

\begin{table}[H]
\centering
\caption{Analytical and extrapolated numerical gradients (hartree/bohr) for \ce{LiH} with the tPBE functional at various Li-H distances (\AA).}
\begin{ruledtabular}
\begin{tabular}{S[table-format=-1.1]S[table-format=-1.10]S[table-format=-1.10]S[table-format=-1.10]S[table-format=-1.10]}
{} & \multicolumn{2}{c}{Analytic} & \multicolumn{2}{c}{Numeric (extrapolated)} \\
{R} & {State 1} & {State 2} & {State 1} & {State 2} \\
\hline
0.4 & -2.6323025249 & -2.6406602924 & -2.6322996605 & -2.6406233188 \\
0.5 & -1.6502107261 & -1.4715468430 & -1.6502084933 & -1.4715506639 \\
0.6 & -1.0957497680 & -0.9447003692 & -1.0956340601 & -0.9447100662 \\
0.7 & -0.7535848001 & -0.6392569040 & -0.7535858442 & -0.6392556906 \\
0.8 & -0.5213096493 & -0.4431537766 & -0.5213127490 & -0.4431523942 \\
0.9 & -0.3599028217 & -0.3117190105 & -0.3598252292 & -0.3117965266 \\
1.0 & -0.2455963991 & -0.2207973687 & -0.2455179608 & -0.2208149582 \\
1.1 & -0.1642346395 & -0.1568320412 & -0.1642313924 & -0.1568321937 \\
1.2 & -0.1063071368 & -0.1114585269 & -0.1062888733 & -0.1114633362 \\
1.3 & -0.0649426063 & -0.0790769996 & -0.0649466146 & -0.0790895671 \\
1.4 & -0.0353596410 & -0.0559404624 & -0.0353574217 & -0.0559432857 \\
1.5 & -0.0142227161 & -0.0394131078 & -0.0142209130 & -0.0394137655 \\
1.6 & 0.0008352634 & -0.0276647306 & 0.0008323424 & -0.0276686996 \\
1.7 & 0.0114870431 & -0.0193440570 & 0.0114916096 & -0.0193425259 \\
1.8 & 0.0189431669 & -0.0135042373 & 0.0189416432 & -0.0135046356 \\
1.9 & 0.0240746537 & -0.0094632936 & 0.0240755643 & -0.0094608638 \\
2.0 & 0.0275187640 & -0.0067475637 & 0.0274997393 & -0.0067283704 \\
2.1 & 0.0296926130 & -0.0048971910 & 0.0297021978 & -0.0049178150 \\
2.2 & 0.0309442241 & -0.0037186214 & 0.0308931184 & -0.0036676157 \\
2.3 & 0.0314896964 & -0.0029654684 & 0.0315066515 & -0.0029916584 \\
2.4 & 0.0315001216 & -0.0025315676 & 0.0315200124 & -0.0025297169 \\
2.5 & 0.0310760534 & -0.0022315616 & 0.0310699606 & -0.0022316258 \\
2.6 & 0.0302977137 & -0.0020104397 & 0.0302982764 & -0.0019459716 \\
2.7 & 0.0292012885 & -0.0017531569 & 0.0289607920 & -0.0014925599 \\
2.8 & 0.0278023179 & -0.0013772033 & 0.0278077622 & -0.0013759877 \\
2.9 & 0.0260689863 & -0.0007775428 & 0.0260509407 & -0.0007778752 \\
3.0 & 0.0239951674 & 0.0001198806 & 0.0239748970 & 0.0001142742 \\
3.1 & 0.0215556777 & 0.0013577489 & 0.0215698253 & 0.0013469852 \\
3.2 & 0.0187647195 & 0.0029343005 & 0.0187454476 & 0.0029052197 \\
3.3 & 0.0157062466 & 0.0047915104 & 0.0157028689 & 0.0047658526 \\
3.4 & 0.0125490719 & 0.0067644819 & 0.0125186591 & 0.0067658219 \\
3.5 & 0.0095215939 & 0.0086252685 & 0.0095875613 & 0.0085955659 \\
3.6 & 0.0068416962 & 0.0101619957 & 0.0068633103 & 0.0101699253 \\
3.7 & 0.0046182478 & 0.0112712975 & 0.0046307996 & 0.0112347406 \\
3.8 & 0.0028629357 & 0.0119527615 & 0.0028225163 & 0.0119728610 \\
3.9 & 0.0015285235 & 0.0122768397 & 0.0015982612 & 0.0123248413 \\
4.0 & 0.0005329732 & 0.0123285726 & 0.0005380740 & 0.0123229700 \\
\end{tabular}
\end{ruledtabular}
\end{table}

\begin{table}[H]
\centering
\caption{Analytical and extrapolated numerical gradients (hartree/bohr) for \ce{LiH} with the ftPBE functional at various Li-H distances (\AA).}
\begin{ruledtabular}
\begin{tabular}{S[table-format=-1.1]S[table-format=-1.10]S[table-format=-1.10]S[table-format=-1.10]S[table-format=-1.10]}
{} & \multicolumn{2}{c}{Analytic} & \multicolumn{2}{c}{Numeric (extrapolated)} \\
{R} & {State 1} & {State 2} & {State 1} & {State 2} \\
\hline
0.4 & -2.6666097780 & -2.6044407224 & -2.6666461205 & -2.6044027729 \\
0.5 & -1.6557892805 & -1.4610902765 & -1.6557879924 & -1.4610812031 \\
0.6 & -1.0921330867 & -0.9464604767 & -1.0921301615 & -0.9464619968 \\
0.7 & -0.7543105174 & -0.6367964921 & -0.7543138708 & -0.6367928945 \\
0.8 & -0.5212120501 & -0.4417074455 & -0.5212099842 & -0.4417070759 \\
0.9 & -0.3598168094 & -0.3106685125 & -0.3598152725 & -0.3106690950 \\
1.0 & -0.2454897428 & -0.2201061723 & -0.2454891993 & -0.2201073348 \\
1.1 & -0.1639980801 & -0.1565294593 & -0.1639970267 & -0.1565290540 \\
1.2 & -0.1064939069 & -0.1109280469 & -0.1064905079 & -0.1109219600 \\
1.3 & -0.0649723061 & -0.0788572212 & -0.0649717814 & -0.0788575570 \\
1.4 & -0.0355543628 & -0.0554964738 & -0.0355546278 & -0.0554961299 \\
1.5 & -0.0144120527 & -0.0388653677 & -0.0144112143 & -0.0388651616 \\
1.6 & 0.0007123451 & -0.0272012251 & 0.0007118142 & -0.0271999382 \\
1.7 & 0.0114814824 & -0.0192846703 & 0.0114837168 & -0.0192608841 \\
1.8 & 0.0188461309 & -0.0131528878 & 0.0188451535 & -0.0131530082 \\
1.9 & 0.0239487456 & -0.0090635354 & 0.0239507087 & -0.0090660696 \\
2.0 & 0.0274335656 & -0.0065705790 & 0.0274337147 & -0.0065708852 \\
2.1 & 0.0295154826 & -0.0044578606 & 0.0295158975 & -0.0044579834 \\
2.2 & 0.0306894116 & -0.0030917634 & 0.0306891583 & -0.0030944659 \\
2.3 & 0.0315608111 & -0.0029551483 & 0.0315616286 & -0.0029561150 \\
2.4 & 0.0319138789 & -0.0026096236 & 0.0319130195 & -0.0026097994 \\
2.5 & 0.0311097457 & -0.0022751738 & 0.0311093458 & -0.0022745924 \\
2.6 & 0.0311781662 & -0.0028443082 & 0.0311795017 & -0.0028473123 \\
2.7 & 0.0297162587 & -0.0019206636 & 0.0297164912 & -0.0019273902 \\
2.8 & 0.0264137990 & 0.0000908253 & 0.0264134477 & 0.0000877646 \\
2.9 & 0.0240125967 & 0.0014145726 & 0.0240088448 & 0.0014153140 \\
3.0 & 0.0257519015 & -0.0011653585 & 0.0257515419 & -0.0011713040 \\
3.1 & 0.0208831991 & 0.0022989715 & 0.0208798270 & 0.0022779257 \\
3.2 & 0.0176145330 & 0.0044678973 & 0.0176179849 & 0.0044492840 \\
3.3 & 0.0148336674 & 0.0059304550 & 0.0148426518 & 0.0058955377 \\
3.4 & 0.0114351802 & 0.0082239968 & 0.0114361708 & 0.0082249193 \\
3.5 & 0.0097880173 & 0.0086775676 & 0.0097898761 & 0.0086686368 \\
3.6 & 0.0064409208 & 0.0108047710 & 0.0064391719 & 0.0108056625 \\
3.7 & 0.0043934033 & 0.0117884700 & 0.0044049851 & 0.0117860007 \\
3.8 & 0.0029413359 & 0.0121522237 & 0.0029448482 & 0.0121455657 \\
3.9 & 0.0033726092 & 0.0108161232 & 0.0033644988 & 0.0108208811 \\
4.0 & 0.0033503427 & 0.0099040769 & 0.0033496967 & 0.0099047238 \\
\end{tabular}
\end{ruledtabular}
\end{table}
